\documentclass[twocolumn,tighten,times]{aastex6}

\usepackage{mathrsfs}
\usepackage{graphicx}
\usepackage{color}

\newcommand{\hb}{\ifmmode {\rm H\beta} \else H$\beta$\fi}
\newcommand{\feii}{\ifmmode {\rm Fe\ II} \else Fe {\sc ii}\fi}
\newcommand{\heii}{\ifmmode {\rm He\ II} \else He {\sc ii}\fi}
\newcommand{\oiii}{\ifmmode {\rm [O\ III]} \else [O {\sc iii}]\fi}
\newcommand{\lag}{\ifmmode {\tau_{\rm H\beta}} \else $\tau_{\rm H\beta}$\fi}
\newcommand{\width}{\ifmmode {V_{\rm FWHM}} \else $V_{\rm FWHM}$\fi}
\newcommand{\mbh}{\ifmmode {M_{\bullet}} \else $M_{\bullet}$\fi}
\newcommand{\fblr}{\ifmmode {f_{\rm BLR}} \else $f_{\rm BLR}$\fi}
\newcommand{\rhb}{\ifmmode {R_{\rm H\beta}} \else $R_{\rm H\beta}$\fi}
\newcommand{\rl}{$R_{\rm H\beta}$--$L_{5100}$}
\newcommand{\dotm}{\ifmmode {\dot{\mathscr{M}}} \else $\dot{\mathscr{M}}$\fi}
\newcommand{\pp}{\ifmmode {^{\prime\prime}} \else $^{\prime\prime}$\fi}
\newcommand{\Rfe}{\ifmmode {{\cal R}_{\rm Fe}} \else ${\cal R}_{\rm Fe}$\fi}
\newcommand{\Dhb}{\ifmmode {{\cal D}_{\hb}} \else ${\cal D}_{\hb}$\fi}

\defcitealias{du2014}{I}
\defcitealias{wang2014}{II}
\defcitealias{hu2015}{III}
\defcitealias{du2015}{IV}
\defcitealias{du2016V}{V}
\defcitealias{du2016VI}{VI}
\defcitealias{xiao2018}{VII}
\defcitealias{li2018}{VIII}

\begin{document}

\title{Supermassive Black Holes with High Accretion Rates in Active Galactic Nuclei. IX \\
10 New Observations of Reverberation Mapping and Shortened \hb\ Lags}

\author{Pu Du\altaffilmark{1,9},
Zhi-Xiang Zhang\altaffilmark{1},
Kai Wang\altaffilmark{1},
Ying-Ke Huang\altaffilmark{1},
Yue Zhang\altaffilmark{1},
Kai-Xing Lu\altaffilmark{2},
Chen Hu\altaffilmark{1},
Yan-Rong Li\altaffilmark{1},
Jin-Ming Bai\altaffilmark{2},
Wei-Hao Bian\altaffilmark{3},
Ye-Fei Yuan\altaffilmark{4},
Luis C. Ho\altaffilmark{5, 6}, and
Jian-Min Wang\altaffilmark{1, 7, 8, 9}\\
(SEAMBH collaboration)}

\altaffiltext{1}{Key Laboratory for Particle Astrophysics, Institute of High Energy Physics,
Chinese Academy of Sciences, 19B Yuquan Road, Beijing 100049, China}

\altaffiltext{2}{Yunnan Observatories, Chinese Academy of Sciences, Kunming 650011, China}

\altaffiltext{3}{Physics Department, Nanjing Normal University, Nanjing 210097, China}

\altaffiltext{4}{Department of Astronomy, University of Science and Technology of China, Hefei 
230026, China}

\altaffiltext{5}{Kavli Institute for Astronomy and Astrophysics, Peking University, Beijing 100871, China}

\altaffiltext{6}{Department of Astronomy, School of Physics, Peking University, Beijing 100871, China}

\altaffiltext{7}{National Astronomical Observatories of China, Chinese Academy of Sciences,
20A Datun Road, Beijing 100020, China}

\altaffiltext{8}{School of Astronomy and Space Science, University of Chinese Academy of Sciences,
19A Yuquan Road, Beijing 100049, China}

\altaffiltext{9}{Corresponding authors: dupu@ihep.ac.cn, wangjm@ihep.ac.cn}

\journalinfo{To appear in {\it The Astrophysical Journal}.}

\begin{abstract}
As one of the series of papers reporting on a large reverberation mapping
campaign of super-Eddington accreting massive black holes (SEAMBHs) in active
galactic nuclei (AGNs), we present the results of 10 SEAMBHs monitored
spectroscopically during 2015--2017. Six of them are observed for the first
time, and have generally higher 5100 \AA\ luminosities than the SEAMBHs
monitored in our campaign from 2012 to 2015; the remaining four are repeat 
observations to check if their previous lags change. Similar to the previous
SEAMBHs, the \hb\ time lags of the newly observed objects are shorter than the
values predicted by the canonical \rl\ relation of sub-Eddington AGNs, by factors
of $\sim2-6$, depending on the accretion rate. The four previously observed objects have lags consistent with previous measurements.
We provide linear regressions for the \rl\ relation,
solely for the SEAMBH sample and for low-accretion AGNs. We find that the relative
strength of \feii\ and the profile of the \hb\ emission line can be used as
proxies of accretion rate, showing that the shortening of \hb\ lags depends
on accretion rates. The recent SDSS-RM discovery of shortened \hb\ lags in
AGNs with low accretion rates provides compelling evidence for retrograde
accretion onto the black hole.
These evidences show that the canonical \rl\
relation holds only in AGNs with moderate accretion rates. At low accretion 
rates, it should be revised 
to include the effects of black hole spin, whereas the accretion rate itself 
becomes a key factor in the regime of high accretion rates.
\end{abstract}


\section{introduction}

Active galactic nuclei (AGNs), powered by accretion onto supermassive black
holes (BHs) in the centers of their host galaxies, are the most luminous and
long-lived sources in the universe.  
The masses of BHs is one of the most critical parameters
controlling the observational properties of AGNs.  Despite significant 
progress in recent years, BH mass estimates in AGNs are still highly 
uncertain.
In the past decades, the reverberation mapping (RM; e.g., \citealt{bahcall1972,
blandford1982, peterson1993}) technique has been demonstrated to be an effective
and efficient way to determine masses of BHs. It measures the delayed response
(\lag) of broad emission lines (e.g., \hb) to the variation of continuum flux.
Combining with the velocity width ($\Delta V$) measured from the full width at half maximum
(FWHM) or the line dispersion $\sigma_{\hb}$ (second moment of the
profile) of the \hb\ emission line,  the BH mass can be simply obtained from
\begin{equation}
\label{eqn:mbh}
\mbh = \fblr\frac{\rhb \Delta V^2}{G},
\end{equation}
where $\rhb = c \lag$ is the emissivity-weighted radius of the broad-line region
(BLR), 
$c$ is the speed of light, $G$ is the gravitational constant, and
\fblr\ is the virial factor that is determined by the geometry, kinematics,
and inclination angle of the BLR. The RM technique has been applied to measure the BH
masses for more than 100 objects by different campaigns
\citep[e.g.,][]{peterson1993, peterson1998, peterson2002, peterson2004,
kaspi2000,  kaspi2007, bentz2008, bentz2009, denney2009, barth2011, barth2013,
barth2015, rafter2011, grier2012, rafter2013, du2014, du2015, du2016V, wang2014,
shen2016, jiang2016, fausnaugh2017, grier2017}. Their results lead to a widely used
relationship between the time delay of \hb\ emission line and the
monochromatic luminosity ($\lambda L_{\lambda}$) at 5100 \AA\  (hereafter
$L_{5100}$).  This canonical \rl\ relation has the form 
\begin{equation}
\label{eqn:r-l_bentz}
\rhb = \alpha \ell^{\beta}_{44},
\end{equation}
where $\ell_{44} = L_{5100} / 10^{44}\ {\rm erg\ s^{-1}}$, and $\alpha$ and $\beta$ are 
constants \citep{kaspi2000, bentz2013}. The \rl\ relation, combined with 
Equation (\ref{eqn:mbh}), has been extensively adopted as a BH mass estimator 
from single-epoch spectroscopy \citep[e.g.,][]{mclure2002, vestergaard2006, 
shen2011,ho2015}. 
However, the canonical \rl\ relation is based mainly on local AGNs of moderate accretion rate and contains only a few objects of high (but not extremely) accretion rate. It does not represent the full range of AGN properties (e.g., accretion rate, BH spin, etc.).

Since 2012, we have been conducting a large RM campaign to monitor
AGNs with high accretion rates.  One of the striking new results of our 
work is that AGNs with high accretion rates deviate significantly from the 
canonical \rl\ relation in exhibiting systematically shorter lags for a 
given luminosity \citep{du2015, du2016V}.  
One of the long-term goals of our campaign is to use super-Eddington
accreting massive black holes (SEAMBHs) as a standard candle
to measure the expansion history of the early Universe \citep{Wang2013}.  
Thus far, our collaboration has published reliable H$\beta$ lags for 20 objects.
Their light curves, cross-correlation
functions (CCFs), and the corresponding time lags have been published in
\cite{du2014, du2015, du2016V}  (hereafter Papers
\citetalias{du2014}, \citetalias{du2015}, and \citetalias{du2016V}), \cite{wang2014} (Paper \citetalias{wang2014}), and
\cite{hu2015} (Paper \citetalias{hu2015}). 
In addition, \citet[Paper \citetalias{du2016VI}]{du2016VI}
analyze the velocity-resolved 
time lags, \citet[Paper \citetalias{xiao2018}]{xiao2018} discuss the velocity-delay maps 
reconstructed by the maximum entropy method \citep{horne1994}, and \citet[Paper \citetalias{li2018}]{li2018} present BH masses measured by BLR dynamical modeling for the objects observed in the first year (from 2012 to 2013).
The \hb\ time lags of those
SEAMBHs are shorter by a factor of 2 -- 8 than the normal AGNs with the same
luminosities (Papers \citetalias{du2015} and \citetalias{du2016V}).  And the lag shortening itself shows strong
correlation with the dimensionless accretion rate, defined as $\dotm =
\dot{M}_{\bullet}/L_{\rm Edd} c^{-2}$, where $\dot{M}_{\bullet}$ is the mass
accretion rate and $L_{\rm Edd}$ is the Eddington luminosity (Papers \citetalias{du2015} and \citetalias{du2016V}).  Thus, we established a new scaling relation, of the form 
\begin{equation}
\rhb=\alpha_1\ell_{44}^{\beta_1}\,\min\left[1,\left(\dotm/\dotm_c\right)^{-\gamma_1}\right],
\end{equation}
where $\alpha_1$, $\beta_1$, and $\gamma_1$ are constants (Paper \citetalias{du2016V}). 
This relation connects the size of the BLR not only with the luminosity but also with the accretion rate.

\begin{table*}
\begin{deluxetable*}{lllclccccr}
\tablecolumns{10}
\tablewidth{\textwidth}
\setlength{\tabcolsep}{6.5pt}
\tablecaption{The sample of SEAMBH2015--2017\label{tab:obj}}
\tabletypesize{\footnotesize}
\tablehead{
\colhead{Object}                      &
\colhead{$\alpha_{2000}$}             &
\colhead{$\delta_{2000}$}             &
\colhead{Redshift}                    &
\colhead{Monitoring Period}           &
\colhead{$N_{\rm spec}$}              &
\colhead{Cadence}                     &
\colhead{}                            &
\multicolumn{2}{c}{Comparison Stars}  \\ \cline{9-10}
\colhead{}                            &
\colhead{}                            &
\colhead{}                            &
\colhead{}                            &
\colhead{}                            &
\colhead{}                            &
\colhead{(days)}                      &
\colhead{}                            &
\colhead{$R_*$}                       &
\colhead{P.A.}                  
}
\startdata
SDSS~J074352.02+271239.5 & 07 43 52.02 & +27 12 39.5 & 0.2520 & 2015 Oct $-$ 2017 Jun & 72 & 5.9 & & $210\pp.4$ & $138^{\circ}.4$ \\
SDSS~J075051.72+245409.3 & 07 50 51.72 & +24 54 09.3 & 0.4004 & 2015 Nov $-$ 2017 May & 61 & 6.4 & & $ 84\pp.0$ & $ 72^{\circ}.6$ \\
SDSS~J075101.42+291419.1 & 07 51 01.42 & +29 14 19.1 & 0.1209 & 2016 Oct $-$ 2017 Jun & 32 & 7.1 & & $133\pp.3$ & $-41^{\circ}.3$ \\
SDSS~J075949.54+320023.8 & 07 59 49.54 & +32 00 23.8 & 0.1879 & 2015 Nov $-$ 2017 Apr & 36 & 7.8 & & $109\pp.2$ & $-48^{\circ}.3$ \\
SDSS~J081441.91+212918.5 & 08 14 41.91 & +21 29 18.5 & 0.1626 & 2016 Oct $-$ 2017 Apr & 24 & 7.0 & & $ 79\pp.0$ & $ 73^{\circ}.9$ \\
SDSS~J083553.46+055317.1 & 08 35 53.46 & +05 53 17.1 & 0.2051 & 2015 Nov $-$ 2017 May & 54 & 6.9 & & $106\pp.3$ & $-42^{\circ}.0$ \\
SDSS~J084533.28+474934.5 & 08 45 33.28 & +47 49 34.5 & 0.3024 & 2016 Oct $-$ 2017 Apr & 27 & 6.5 & & $205\pp.5$ & $-126^{\circ}.4$ \\
SDSS~J093302.68+385228.0 & 09 33 02.68 & +38 52 28.0 & 0.1772 & 2016 Oct $-$ 2017 Jun & 65 & 3.5 & & $ 57\pp.7$ & $-156^{\circ}.2$ \\
SDSS~J100402.61+285535.3 & 10 04 02.61 & +28 55 35.3 & 0.3272 & 2015 Nov $-$ 2017 Jun & 89 & 4.9 & & $ 75\pp.6$ & $ 41^{\circ}.8$ \\
SDSS~J101000.68+300321.5 & 10 10 00.68 & +30 03 21.5 & 0.2564 & 2015 Nov $-$ 2017 Jun & 70 & 5.9 & & $163\pp.4$ & $-96^{\circ}.7$ 
\enddata
\tablecomments{$N_{\rm spec}$ is the number of spectroscopic epochs. $R_*$ is the angular distance between the object 
and the comparison star. P.A. is the position angle of the comparison star from the object. 
``Cadence'' is the average sampling interval of the objects.
}
\end{deluxetable*}
\end{table*}

The 5100 \AA\ luminosities of the SEAMBHs observed between 2012 October and 2015 June
(hereafter SEAMBH2012--2014) range from $10^{43}$ to $10^{44.5}\ {\rm
erg\ s^{-1}}$. In contrast, the luminosities of the RM AGNs with normal accretion rates 
span $10^{41.5}$ to $10^{46}\ {\rm erg\ s^{-1}}$ (see Figure 2
in Paper \citetalias{du2016V}). In order to improve the completeness of the
SEAMBH sample,  it is necessary to observe more SEAMBHs with higher and lower
luminosities. From 2015 October to 2017 June, we monitored six SEAMBHs with
luminosities $L_{5100} = 10^{44}-10^{45.5}\ {\rm erg\ s^{-1}}$, 
generally more powerful than the objects in SEAMBH2012--2014. Besides, we also observed
four objects in SEAMBH2012--2014 that had relatively poorer \hb\ lag measurements (e.g.,
the scatter of their light curves is relatively larger, or the length of
their light curves is relatively shorter) than the other sources covered in 
the campaign. We observed them again in order to confirm their \hb\ time
lag measurements.  The coordinates and some other information of the objects
are listed in Table \ref{tab:obj}.

In this paper, we report the results of the SEAMBHs observed during 2015 October --
2017 June (hereafter SEAMBH2015--2016). The target selection, observation, and
data reduction are described in Section \ref{sec:observation}. The light
curves, the lag measurements, and their BH masses and accretion rates are
provided in Section \ref{sec:lag_measurement},  along with notes for each
individual object.  Their positions in the \rl\ relation are shown in Section
\ref{sec:r-l}.  Some discussions are provided in Section \ref{sec:discussion}.
And we give a short summary in Section \ref{sec:summary}. In this work, as in 
other papers in this series, we use a standard 
$\Lambda$CDM cosmology with $H_0=67~{\rm km~s^{-1}~Mpc^{-1}}$, $\Omega_{\Lambda}=0.68$, and 
$\Omega_m=0.32$ \citep{ade2014}.

\section{Observation and Data reduction}
\label{sec:observation}

The details of the target selection, telescope, instrument, observation, and
data reduction of the SEAMBH2015--2016 campaign are, with only minor exceptions, almost the same as those for the
observations in 2013 October -- 2015 June (hereafter SEAMBH2013--2014, Papers
\citetalias{du2015} and \citetalias{du2016V}). In
this section, we introduce the differences from SEAMBH2013--2014 and
briefly summarize the same points for completeness.

\subsection{Target Selection}
\label{sec:target}

Similar to SEAMBH2013--2014 (Papers
\citetalias{du2015} and \citetalias{du2016V}), we selected SEAMBH candidates
based on the dimensionless accretion rate estimator derived from the standard
thin accretion disk model \citep{shakura1973}. From the standard disk model,
the dimensionless accretion rate is given by (see more details in Paper
\citetalias{wang2014} and Appendix A in Paper \citetalias{du2016V})
\begin{equation}
\label{eqn:mdot}
\dotm = 20.1 \left(\frac{\ell_{44}}{\cos i}\right)^{3/2} m^{-2}_7,
\end{equation}
where $m_7 = \mbh / 10^7 M_{\odot}$, and $i$ is inclination angle of disk to
the line of sight. We took $\cos i = 0.75$ (see some discussions in Paper
\citetalias{du2016V}), which is an average estimate for type I AGNs (e.g.,
\citealt{fischer2014, pancoast2014}). For BH masses, we adopted the virial
factor $\fblr = 1$ in our series of papers (see more discussions in 
Section \ref{sec:mass_mdot} and in Paper \citetalias{du2015}).

In SEAMBH2013--2014, we fitted the spectra of all the quasars in Data Release
7 of Sloan Digital Sky Survey (SDSS) using the fitting procedure in
\cite{hu2008a, hu2008b}, then estimated their BH masses and accretion rates by
applying the normal \rl\ relation in \cite{bentz2013}. However, high--$\dotm$
objects tend to have shortened \hb\ time lags (Papers \citetalias{du2015} and
\citetalias{du2016V}), which means the normal \rl\ relation may underestimate
their accretion rates. \cite{du2016F} discovered a bivariate correlation between
\dotm\ and the profile of broad \hb\ line ($\Dhb = {\rm FWHM}/\sigma_{\hb}$),
and the flux ratio of optical \feii\ to \hb\ (\Rfe).
This correlation provides a straightforward method to determine
$\dotm$ from single-epoch spectra of AGNs, and applies to a wide range of
accretion rates ($\dotm \approx 10^{-2} - 10^3$). We can easily estimate
the accretion rates after we measure \Rfe\ and \Dhb\ from the spectra of SDSS
quasars by multi-component fitting procedure in \cite{hu2008a, hu2008b}.
Therefore, instead of the accretion rates estimated from the traditional \rl\
relation and Equation (\ref{eqn:mdot}), we adopted the \dotm\ estimated from \Rfe\ and \Dhb\ to
choose the targets in SEAMBH2015--2016.

After measuring \dotm\ for all of the SDSS quasars, we selected the objects
with the highest \dotm\ and $L_{5100} \approx 10^{44} - 10^{45.5}\ {\rm erg\ s^{-1}}$.
The coordinates of the targets should be appropriate for the site  of the
observatory. Furthermore, we constrained the redshift ($z \approx 0.2-0.4$), and the SDSS $r$-band 
magnitude ($r^{\prime}<17.5$) in order to
obtain high enough signal-to-noise ratio (S/N). At the beginning, 10 objects were chosen as the
observing targets. However, four of them were observed only in the first $\sim$2
months (with only a few epochs) because of poorer weather and the very limited
observing time; these were rejected thereafter. The remaining six objects were
observed for entirely two years.   Besides the six objects with relatively
high luminosities, we monitored four SEAMBHs that had been observed previously
in SEAMBH2013--2014. The uncertainties of their  \hb\ lag measurements were
larger than those of the other objects in SEAMBH2013--2014 because their
light curves were shorter or the scatter of the points in the light
curves was relatively larger. We observed them again in 2015 -- 2017, in order
to check their former measurements.  In total, the 10 objects listed in Table
\ref{tab:obj} constitute the sample in the present paper.

\begin{table}
\begin{deluxetable}{cccc}
\tablecolumns{4}
\tablewidth{\textwidth}
\setlength{\tabcolsep}{6pt}
\tablecaption{Continuum and \hb\ windows in the rest frame\label{tab:window}}
\tabletypesize{\footnotesize}
\tablehead{
\colhead{Object}           &
\colhead{Continuum (blue)}  &
\colhead{\hb}              &
\colhead{Continnum (red)}   \\  
\colhead{}                 &
\colhead{(\AA)}            &
\colhead{(\AA)}            &
\colhead{(\AA)}               
}
\startdata
SDSS~J074352 & 4740--4780 & 4810--4910 & 5075--5125 \\
SDSS~J075051 & 4740--4790 & 4810--4910 & 5075--5125 \\
SDSS~J075101 & 4740--4790 & 4810--4910 & 5075--5125 \\
SDSS~J075949 & 4740--4790 & 4810--4910 & 5075--5125 \\
SDSS~J081441 & 4750--4790 & 4810--4910 & 5075--5125 \\
SDSS~J083553 & 4740--4780 & 4800--4900 & 5075--5125 \\
SDSS~J084533 & 4740--4790 & 4810--4910 & 5075--5125 \\
SDSS~J093302 & 4750--4790 & 4810--4910 & 5075--5125 \\
SDSS~J100402 & 4750--4790 & 4810--4910 & 5075--5125 \\
SDSS~J101000 & 4740--4780 & 4810--4910 & 5075--5125 
\enddata
\end{deluxetable}
\end{table}

\subsection{Photometry and Spectroscopy}
\label{sec:obs}

The photometric and spectroscopic data used in this work were taken with the
Lijiang 2.4 m telescope at the Yunnan Observatories of the Chinese Academy of
Sciences. The telescope is equipped with the Yunnan Faint Object Spectrograph
and Camera,  which is a multi-functional instrument available both for
photometry and spectroscopy.  The images were taken using the
SDSS $r^{\prime}$ filter. For spectroscopy, we used  Grism 3, which has a 
sampling of 2.9 \AA\ pixel$^{-1}$ ($\sim$108 km s$^{-1}$ pixel$^{-1}$) and a
wavelength coverage of 3800 -- 9000 \AA. To minimize the influence from 
atmospheric differential refraction, we employed a longslit
with a width of 5\arcsec. For each object, we oriented the slit to observe 
simultaneously a nearby comparison star\footnote{For SDSS J093302, we used a comparison star
($R_{*}=184\pp.6$; P.A.=$284^{\circ}.3$) different from the one listed in
Table \ref{tab:obj} before 2016 Dec 12. We discovered that our previous observations were 
adversely affected by another very bright star in the slit, located 
between the target and our previous comparison star, which
contributed a very high background to both the target and the
comparison star.  This reduced the S/N of the previous spectra and 
increased the scatter of the 
light curves. After changing
the comparison star to the one listed in Table \ref{tab:obj},
the quality of the new light curves, especially the continuum,
is relatively better.} (listed in Table \ref{tab:obj}) as a calibration standard.
This method provides
highly accurate flux calibration (see Papers \citetalias{du2014} --
\citetalias{du2016V}).

Both the photometric and spectroscopic data were reduced with IRAF v2.16.
The photometric light curves of the targets and the comparison stars were
generated by differential photometry using several (5--8) other stars 
in the same fields. The radius of the aperture used for photometry is typically
4\arcsec, and the annulus for background determination is 8\arcsec.5 to 17\arcsec. The
comparison stars  themselves are very stable given the photometric light
curves shown in Figure \ref{fig:comp} in Appendix \ref{sec:comp}, and thus they can be
used as standards for the spectral calibration.

The spectra were extracted using a uniform aperture of 8\arcsec.5 and a
background region of 7\arcsec.4 -- 14\arcsec on both sides of the aperture. The
fiducial spectra of the comparison stars were produced using data from 
nights with photometric conditions. The fluxes of the target spectra were calibrated
by the comparison stars in the slit. More details for the photometry and
spectroscopy can be found in Papers \citetalias{du2015} and
\citetalias{du2016V}.

The \oiii-based calibration approach \citep{vanGroningen1992,
fausnaugh2017a} is widely used in many RM works \cite[e.g.,][]{peterson1998,
bentz2009, grier2012, fausnaugh2017}, but it does not apply to the SEAMBHs
(Papers \citetalias{du2014}, \citetalias{du2015}, and \citetalias{du2016V}). The \oiii\ $\lambda$5007 emission
lines in SEAMBHs are too weak to be used as a standard for flux
calibration (see the mean spectra in Papers \citetalias{du2015} and \citetalias{du2016V}).
Even worse, the \feii\ contribution beneath the \oiii\ line, which is shown to
also reverberate (\citealt{barth2013}, Paper \citetalias{hu2015}), is relatively
strong. Thus, the \oiii-based calibration approach \citep{vanGroningen1992,
fausnaugh2017a} may give rise to large uncertainties in the flux calibration
of the SEAMBHs in this paper. It has been demonstrated that the method based
on the comparison star can provide accurate flux calibration: for
the SEAMBHs with moderate \oiii\ (in Paper \citetalias{du2014}),  the
variation of the \oiii\ fluxes in the calibrated spectra (by the comparison
stars) is $\sim$3\% (Paper \citetalias{du2014}); for low-accretion rate AGNs (e.g.,
NGC 5548), the \oiii\ fluctuation after the calibration is at a level of
2\% \citep{lu2016}. Therefore, we adopt the calibration approach based on 
comparison star as in Papers \citetalias{du2014}-\citetalias{du2016V}. In Appendix \ref{sec:calib}, as an example, we show that 
the scatter of \oiii\ fluxes in the calibrated spectra of SDSS J075101 is $\lesssim3\%$, which can be regarded as  
an estimate for the calibration precision in this paper.

\begin{figure*}
\centering
\includegraphics[width=0.75\textwidth]{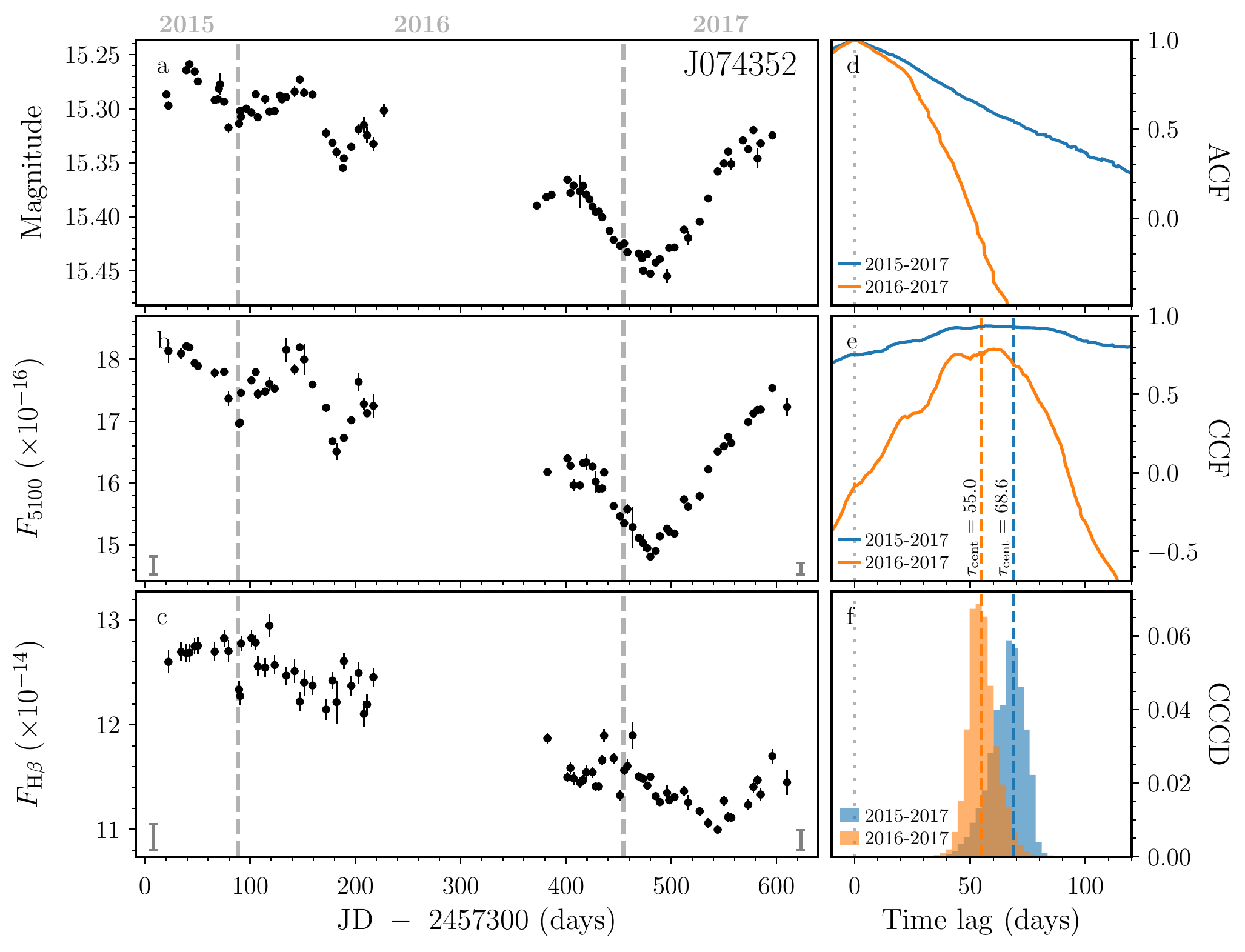} \\
\includegraphics[width=0.75\textwidth]{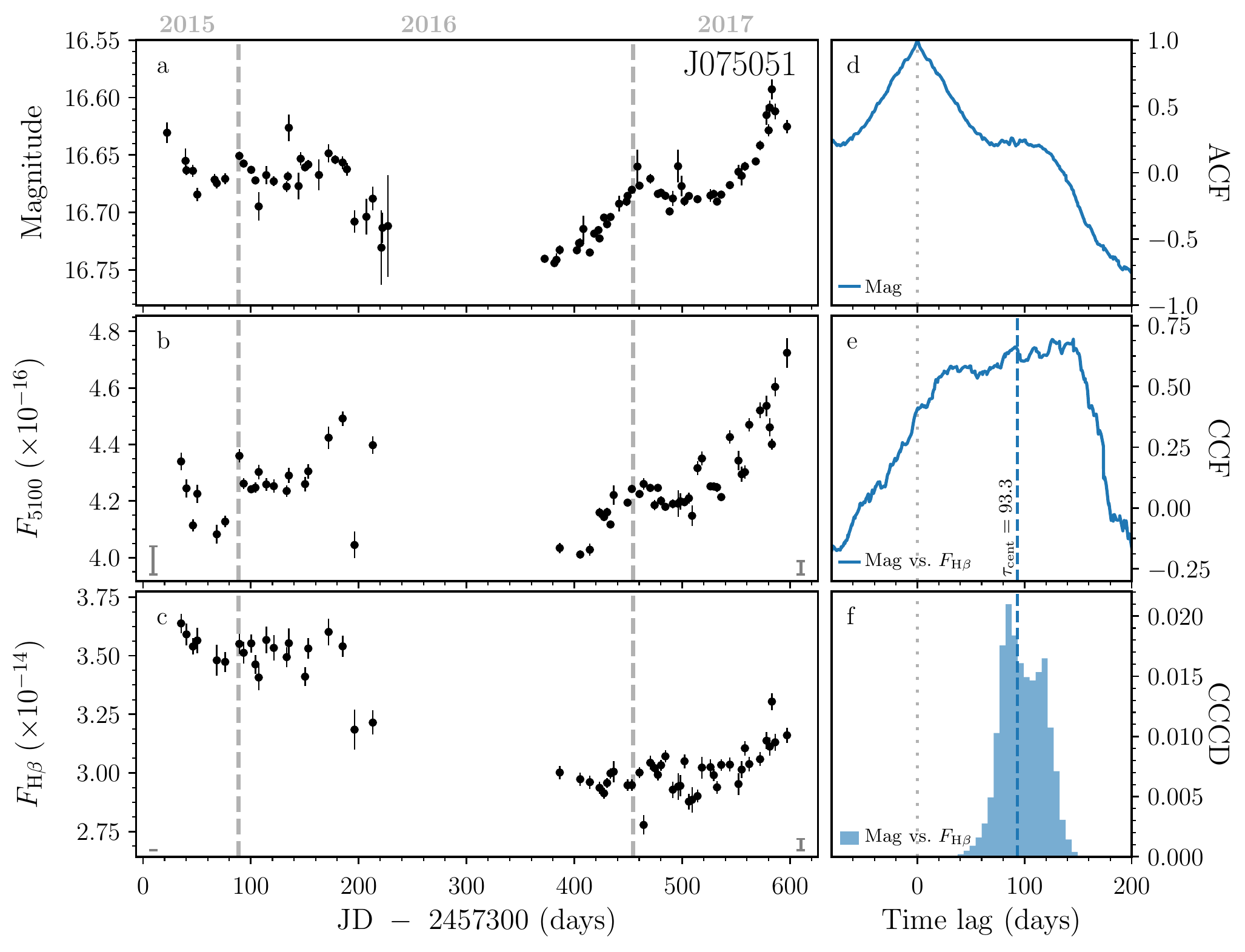} 
\caption{Light curves and cross-correlation functions. Panels a, b, c are the
photometric, 5100 \AA\ continuum, and \hb\ light curves. The name of the object is marked on the upper-right corner in panel a.
The gray error bars in
the lower-left and lower-right corners are the systematic errors in 2015 --
2016 and in 2016 -- 2017, respectively (see more details in Section
\ref{sec:light_curves}). The units of the continuum and emission-line light curves are
${\rm erg\ s^{-1}\ cm^{-2}\ \AA^{-1}}$ and ${\rm erg\ s^{-1}\ cm^{-2}}$, respectively.
The gray dashed lines in panels a, b, c mark the beginnings (January 1st) of the different years.  
Panels d, e, f are the ACF, CCF, and the cross-correlation
 centroid distribution (CCCD) in the observed frame. The orange and blue colors in panels
d, e, f indicate that we perform CCF analysis by using the entire light curves
or the light curves in a single year. The gray dotted lines mark the zero time lags. The time lags in the observed frame are marked by the dashed lines (with the numbers beside) in panels e and f.
For SDSS~J075051, the time lag is obtained by the CCF of the photometric and \hb\ light 
curves. For SDSS~J075949, we also plot its light curves in 2014--2015 from Paper \citetalias{du2016V}
for comparison. }
\label{fig:light_curves}
\end{figure*}

\begin{figure*}
\figurenum{\ref{fig:light_curves}}
\centering
\includegraphics[width=0.75\textwidth]{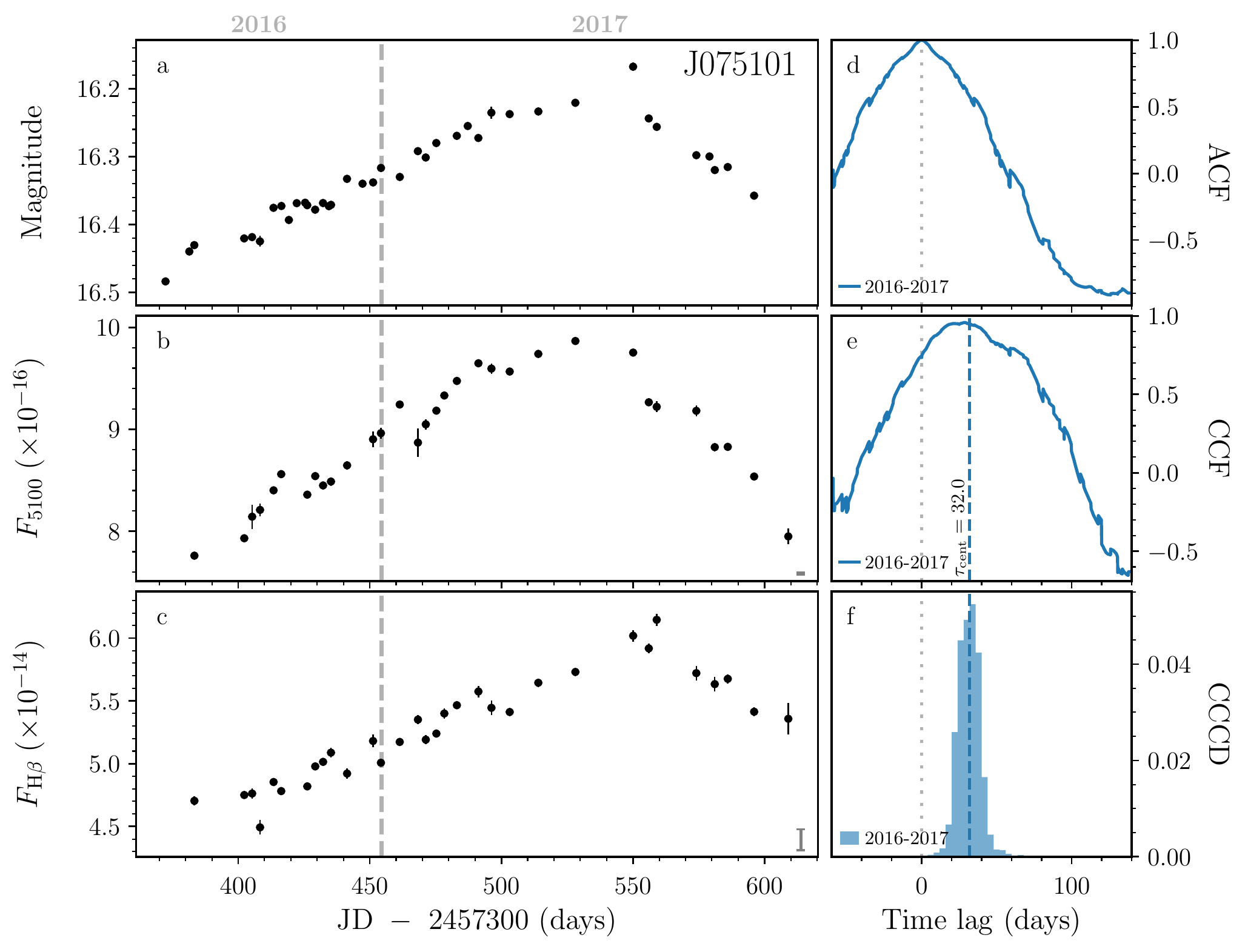} \\
\includegraphics[width=0.75\textwidth]{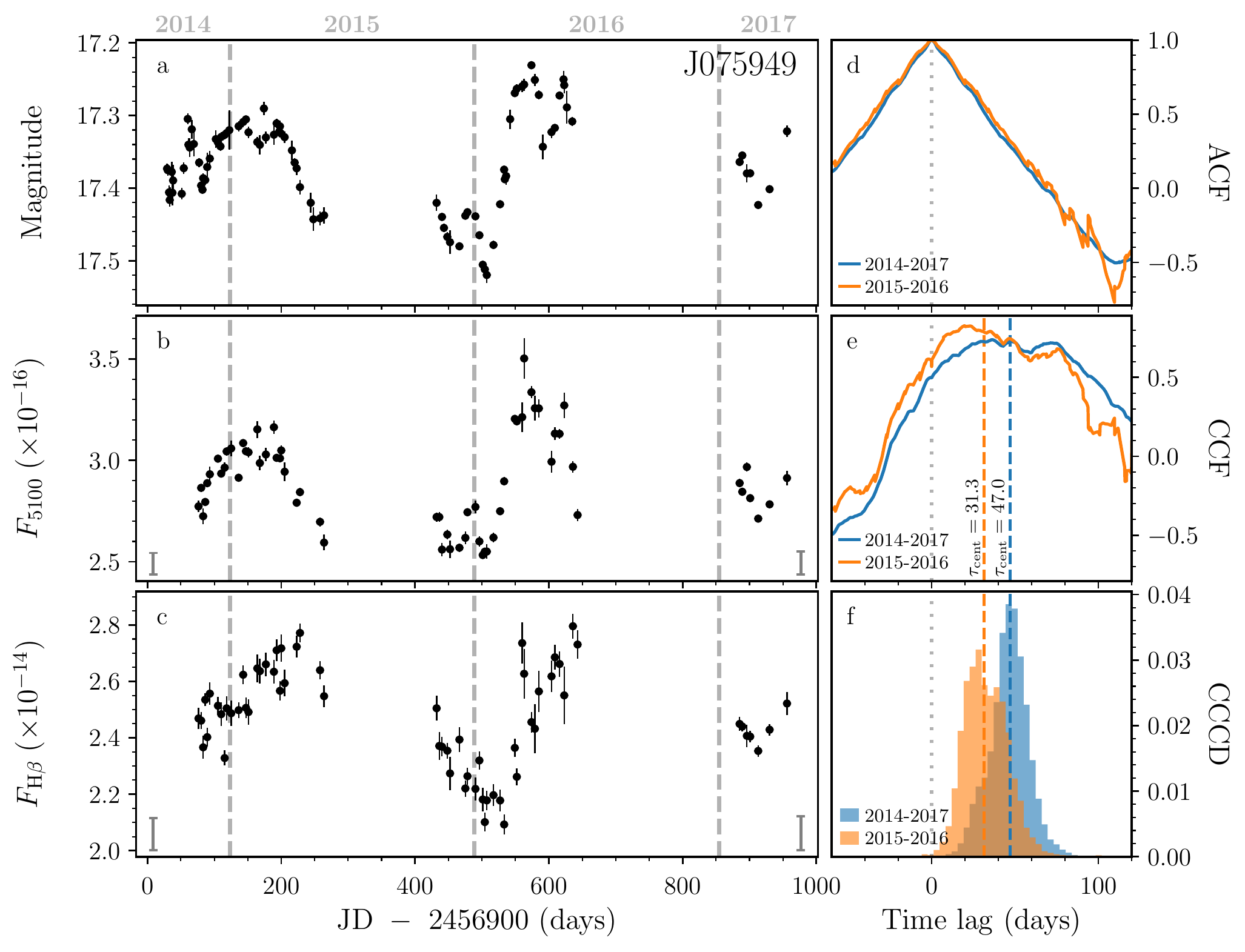} 
\caption{(Continued.)}
\end{figure*}

\begin{figure*}
\figurenum{\ref{fig:light_curves}}
\centering
\includegraphics[width=0.75\textwidth]{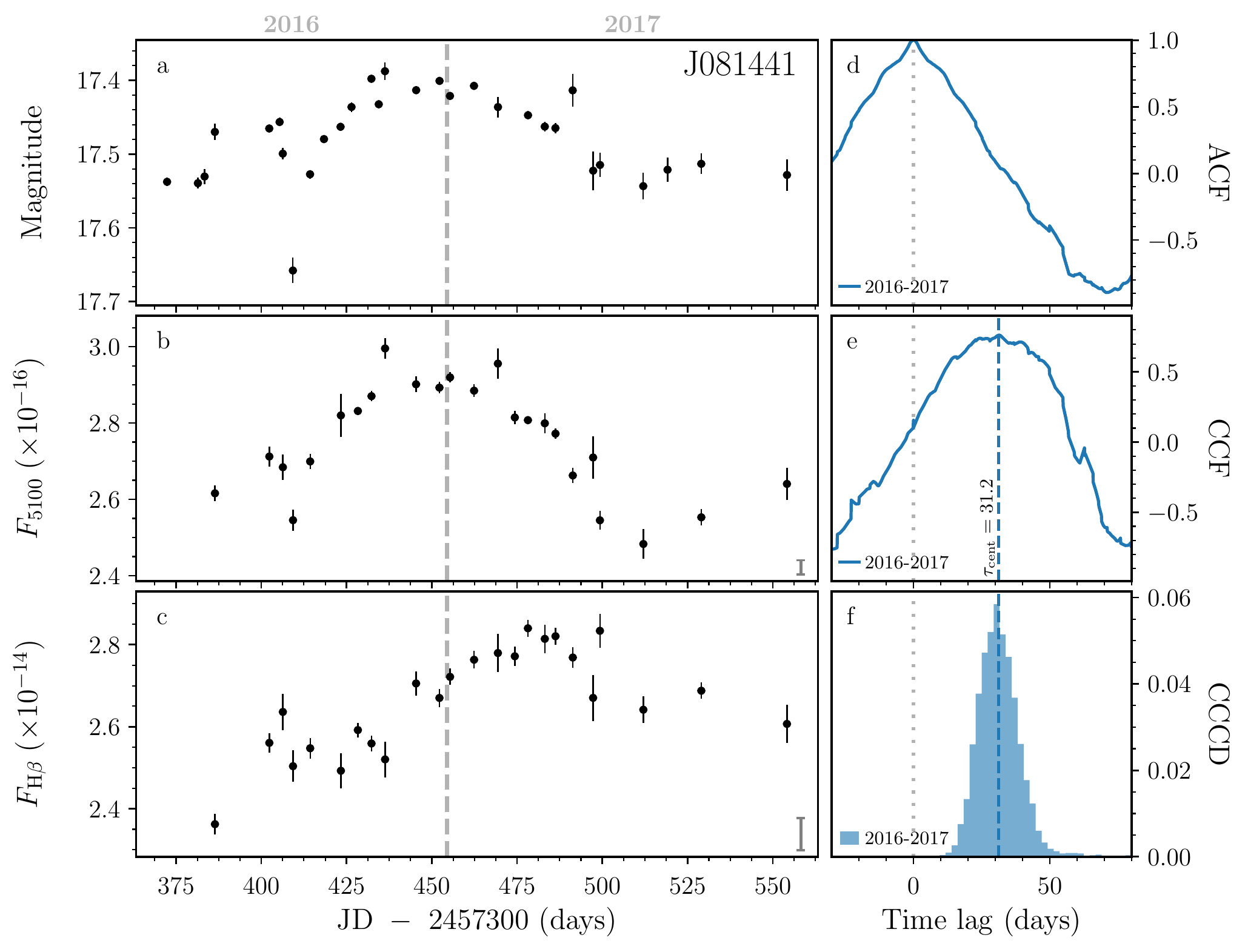} \\
\includegraphics[width=0.75\textwidth]{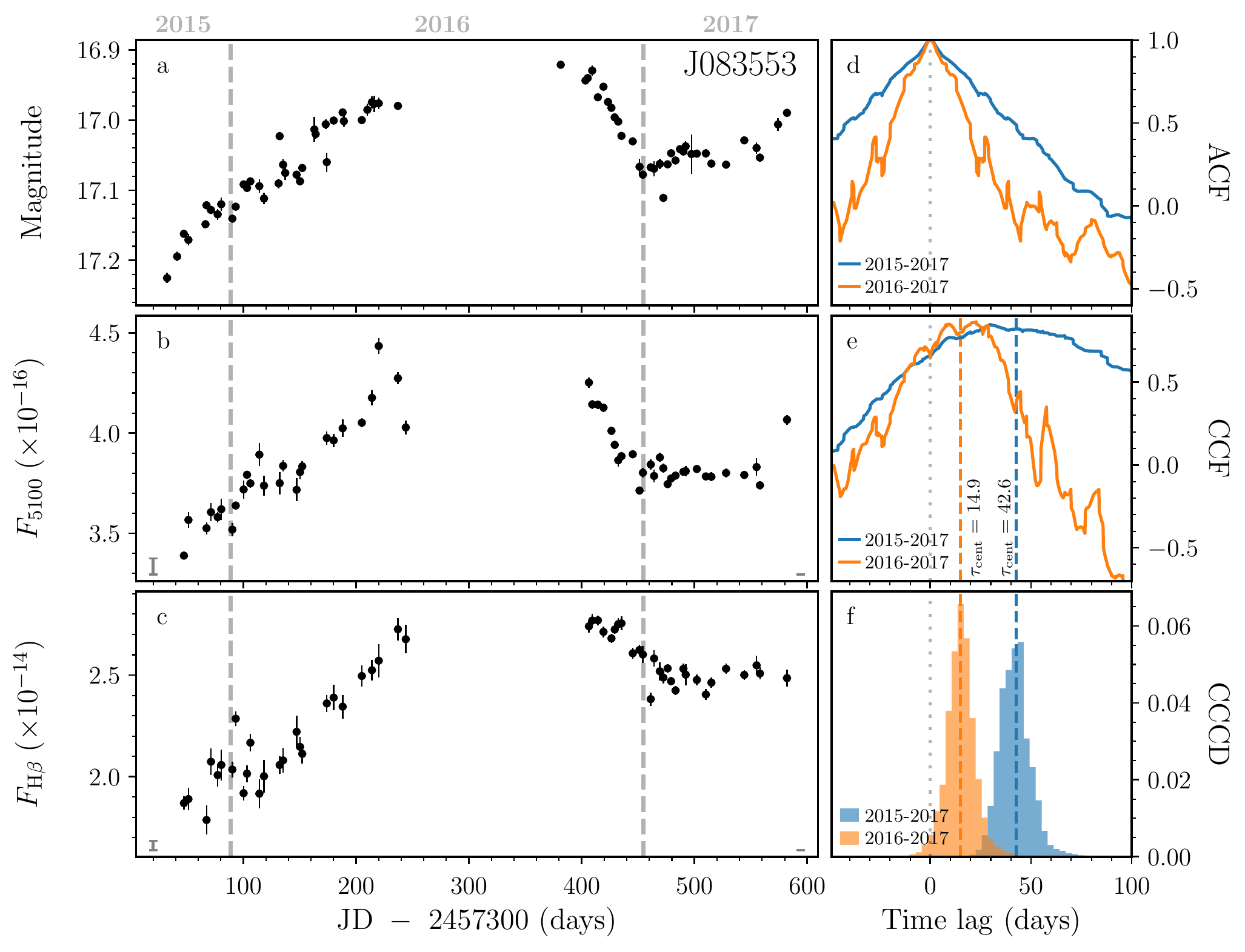} 
\caption{(Continued.)}
\end{figure*}

\begin{figure*}
\figurenum{\ref{fig:light_curves}}
\centering
\includegraphics[width=0.75\textwidth]{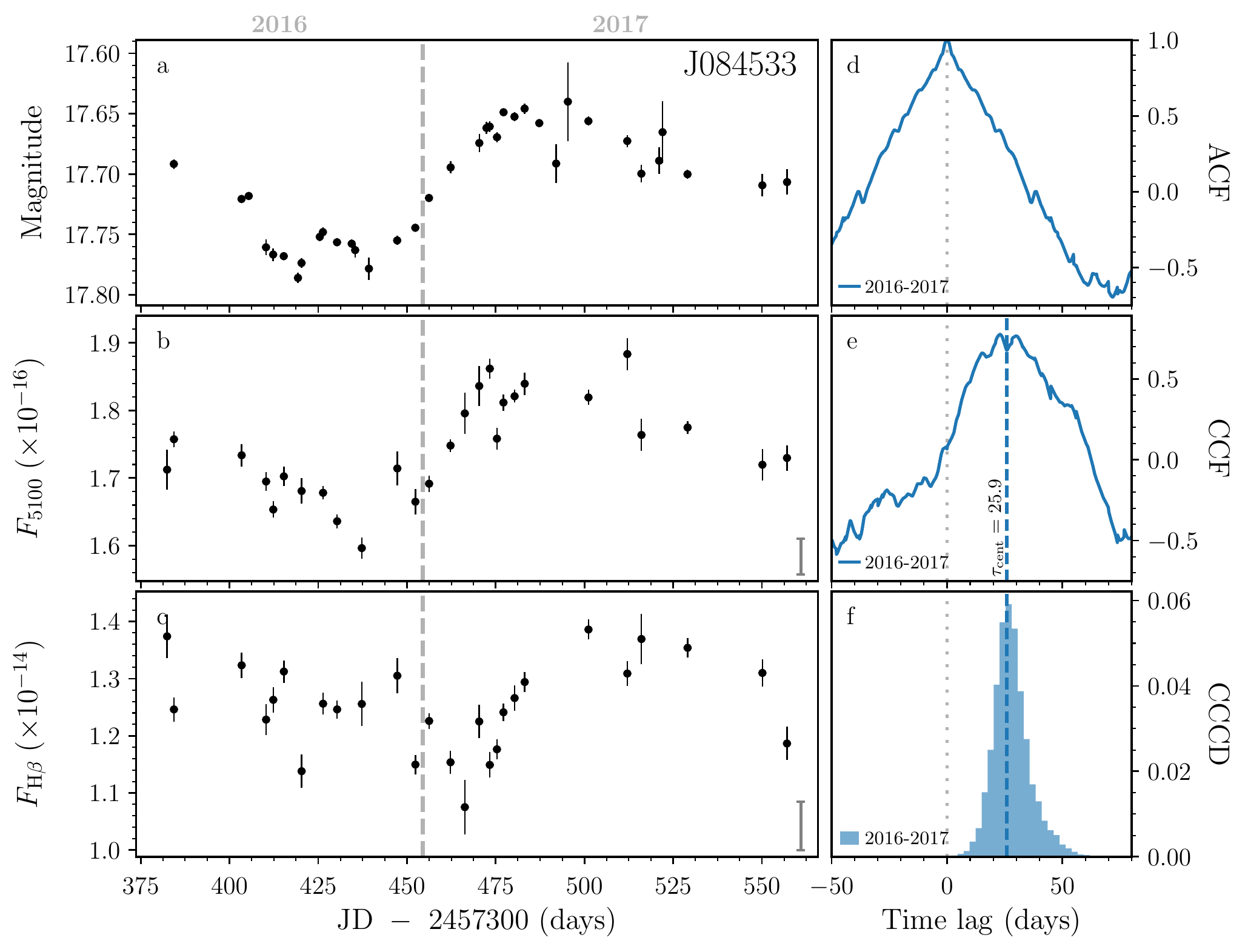} \\
\includegraphics[width=0.75\textwidth]{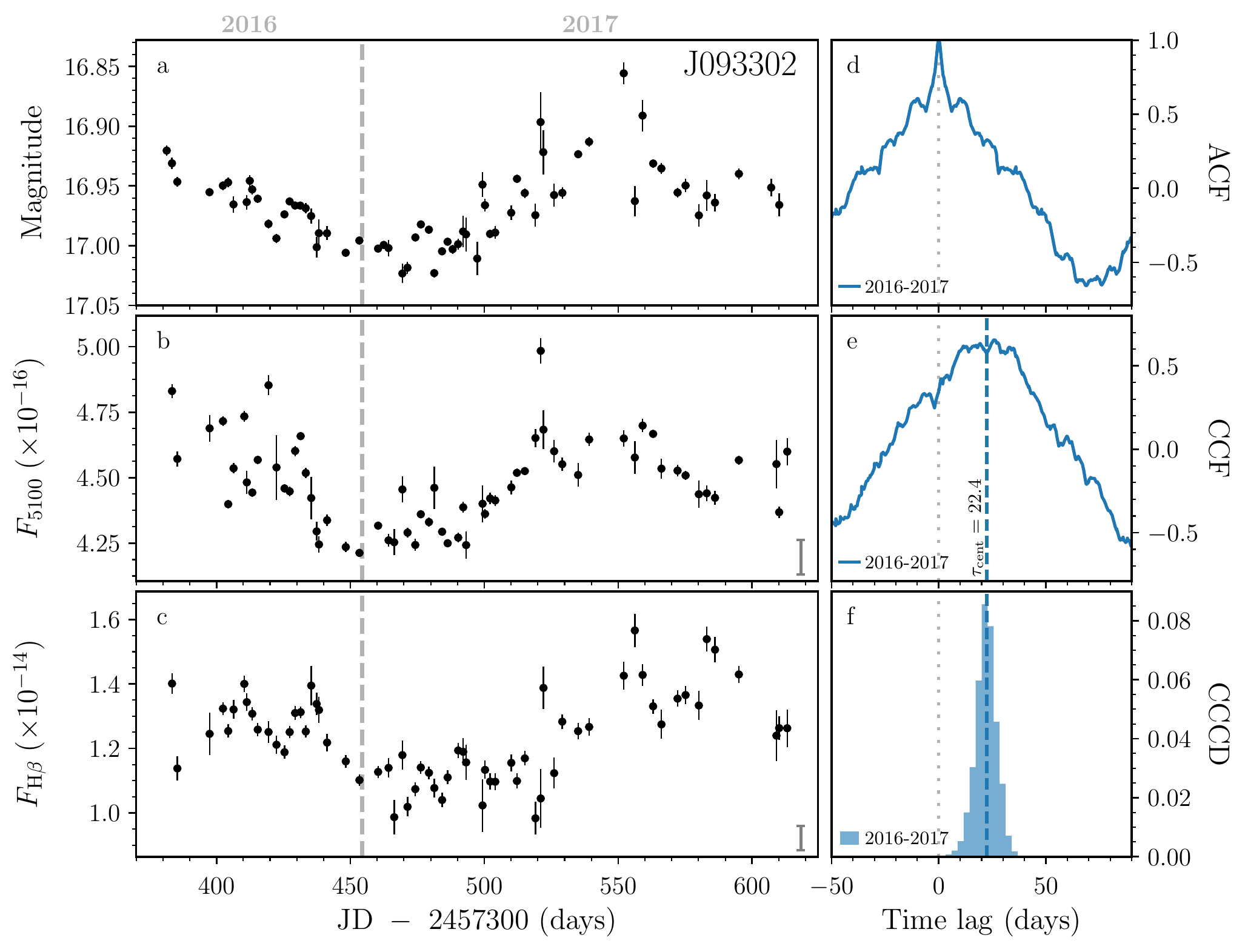} 
\caption{(Continued.)}
\end{figure*}

\begin{figure*}
\figurenum{\ref{fig:light_curves}}
\centering
\includegraphics[width=0.75\textwidth]{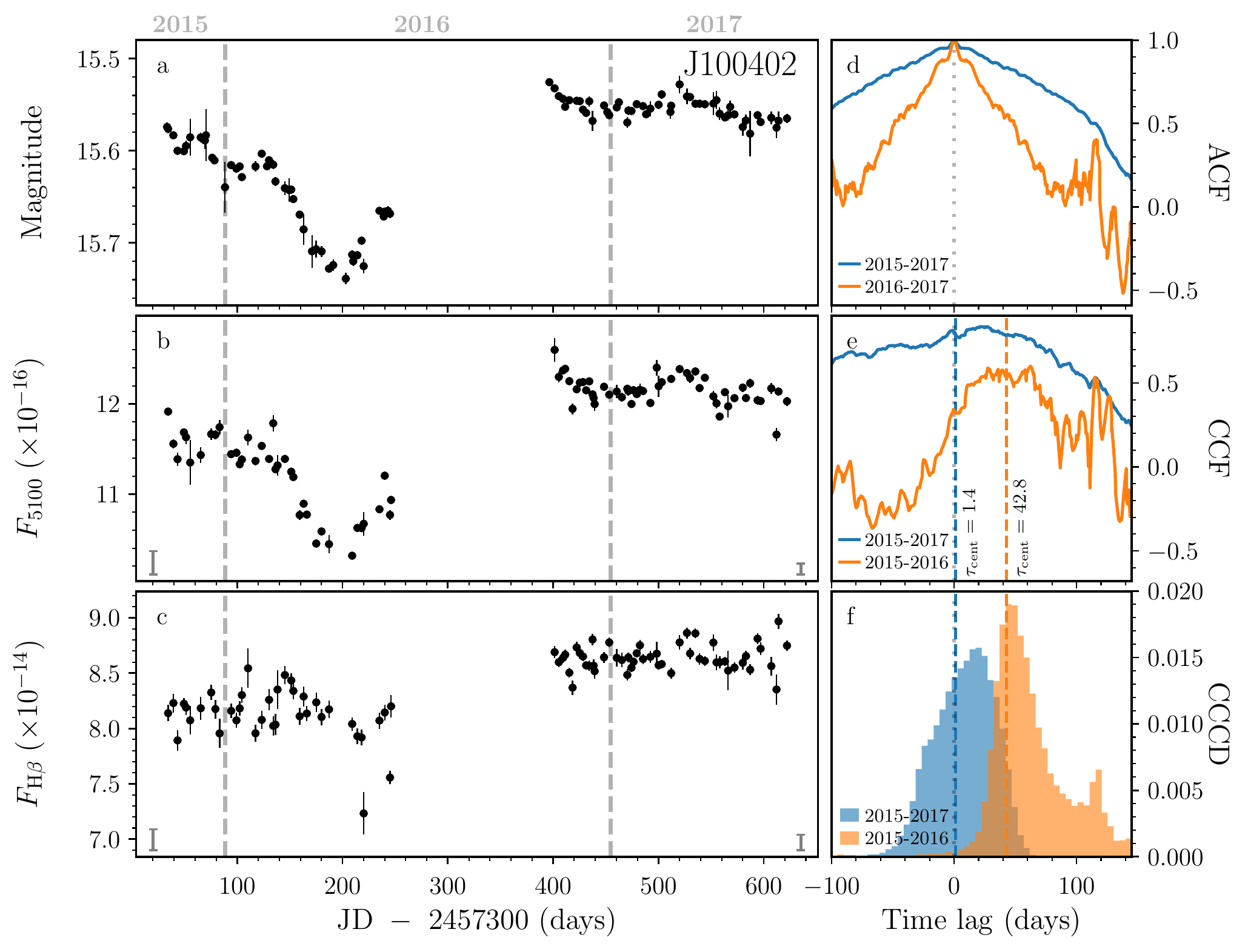} \\
\includegraphics[width=0.75\textwidth]{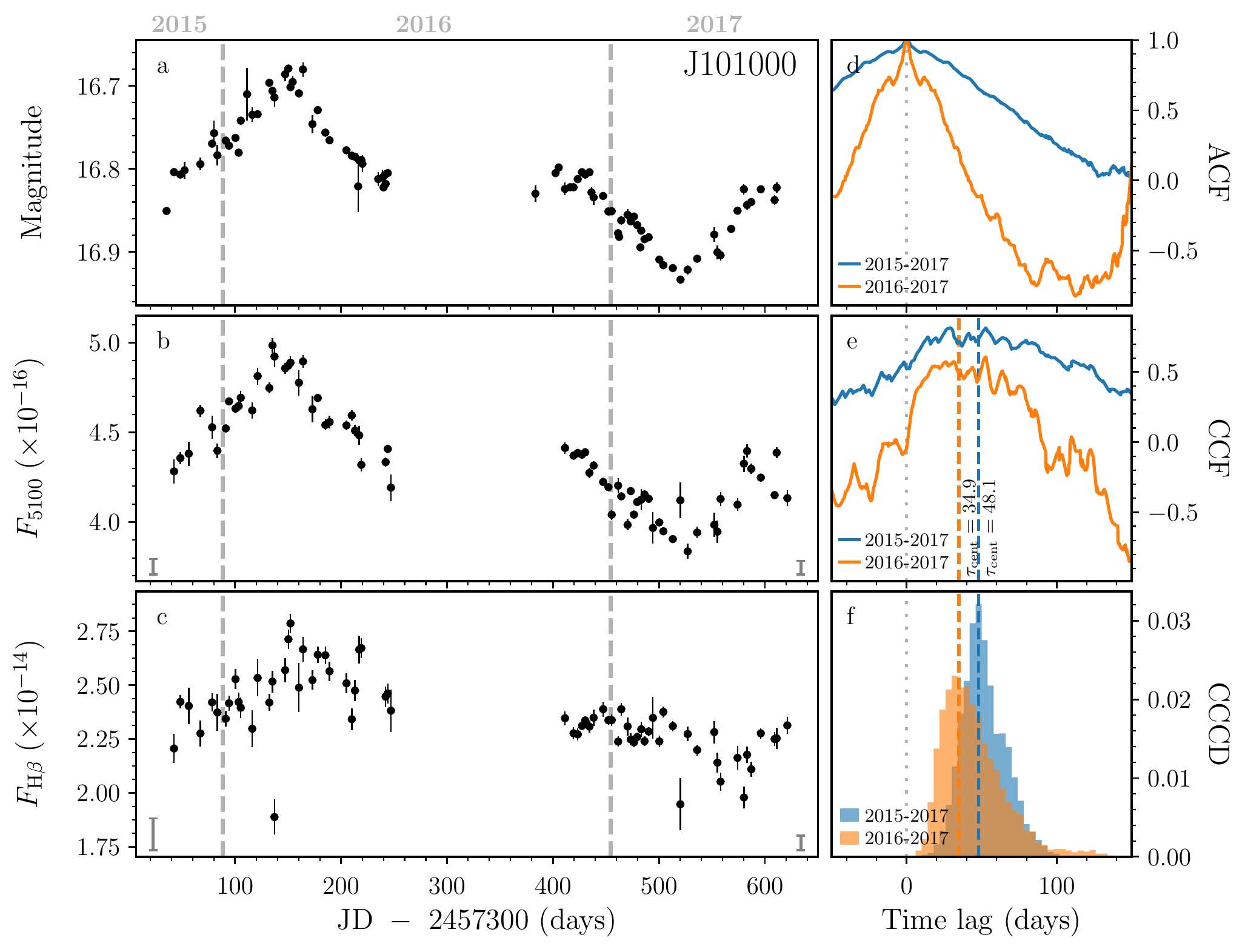} 
\caption{(Continued.)}
\end{figure*}

\section{Measurements of time lags, black hole masses, and accretion rates}
\label{sec:lag_measurement}

\subsection{Light Curves}
\label{sec:light_curves}

After the data reduction and calibration, we can measure their continuum and
\hb\ light curves from the calibrated spectra.  The light curves 
can be obtained by two different approaches: (1) direct
integration method and (2) spectral decomposition method. The first approach,
widely used in most RM studies (e.g., \citealt{peterson1998, kaspi2000, bentz2009, grier2012, fausnaugh2017}, Papers \citetalias{du2014}, \citetalias{du2015}, and \citetalias{du2016V}), simply integrates the flux in the \hb\ band after subtracting the continuum determined by two nearby line-free windows.
It applies to strong and isolated emission
lines (e.g., \hb) and works well both for the spectra with good or poor S/N. The
second one measures the emission-line fluxes by multi-component spectral
fitting, and has been gradually adopted in recent years
(e.g., \citealt{barth2013, barth2015}, Paper \citetalias{hu2015}, and \citealt{lu2016}). It can deblend and
measure the fluxes of \feii\ or \heii\ lines, but has higher demand for the S/N of the spectra. The
spectral decomposition approach can remove the contamination from \feii\ in the \hb\ flux
measurements. However, given the better robustness of the integration method
for spectra with moderate S/N, we adopt the direct
integration approach, as a first step, to measure the continuum and \hb\ light
curves in this paper. It should be noted that the two approaches do not give
very different \hb\ lag measurements for the AGNs with high accretion rates
(see Figure 6 in Paper \citetalias{hu2015}). We will measure their \feii\ light curves 
using the spectral decomposition method in a separate paper in the future.

We choose the continuum and the
\hb\ windows that can  avoid the contamination from the other emission lines
(e.g., \oiii$\lambda$4959, \feii, and \heii) as much as possible for each
object (listed in Table \ref{tab:window}).  The continuum fluxes are set as
the median values in the windows around 5100 \AA\ in the rest frame.  To
measure the line light curves, we integrate the fluxes in the windows of \hb\
after subtracting the background beneath, which is determined by interpolating
two nearby bands (blue and red continuum windows, see also in Figure
\ref{fig:meanrms} in Appendix \ref{sec:meanrms}).  
The uncertainties in the
light curves consist of two components: statistical noises originated from the
Poisson process of photons and systematic uncertainties caused by poor weather
conditions, bright moon, telescope tracking inaccuracies, slit positioning,
etc. The Poisson noises are demonstrated as the error bars of the points in
the light curves (see Figure \ref{fig:light_curves}). The systematic
uncertainties are estimated  by the median filter method (see more details in
Paper \citetalias{du2014}), and are marked as gray error bars in the lower-left
 corners (for the light curves taken in 2015 Oct -- 2016 Jun) and  lower-right
  corners (for the light curves taken in 2016 Oct -- 2017
Jun)\footnote{For SDSS~J075949, the error bars marked in the lower-left corners
are the systematic uncertainties for the light curves in 2014-2015, and the
error bars in the lower-right corners are the systematic uncertainties for
2015-2017.} of panels b and c in Figure \ref{fig:light_curves}. Both the
two components of the uncertainties are taken into account in the analysis of 
the following sections. The
photometric light curves of the targets, the continuum light curves at
5100 \AA\, and \hb\ light curves are provided in Tables
\ref{tab:074352_075051}-\ref{tab:100402_101000} and shown in Figure
\ref{fig:light_curves}.
We plot the photometric light curves to verify the calibration
precision of the spectra. It is obvious that the calibration based on the comparison stars works fine,
because the $r^{\prime}$-band light curves are consistent with the 5100 \AA\ light curves (Figure \ref{fig:light_curves}).
In addition, the photometric light curves can also be used as a substitute of the continuum, 
if the quality of the 5100 \AA\ light curve is not good enough (see Section \ref{sec:individual_obj}).

\begin{table*}
\begin{deluxetable*}{rlcrllcrlcrll}
  \tablecolumns{13}
  \setlength{\tabcolsep}{5pt}
  \tablewidth{0pc}
  \tablecaption{Light curves of SDSS~J074352 and SDSS~J075051\label{tab:074352_075051}}
  \tabletypesize{\scriptsize}
  \tablehead{
      \multicolumn{6}{c}{SDSS~J074352}        &
      \colhead{}                         &
      \multicolumn{6}{c}{SDSS~J075051}        \\ \cline{1-6}\cline{8-13}
      \multicolumn{2}{c}{Photometry}     &
      \colhead{}                         &
      \multicolumn{3}{c}{Spectra}        &
      \colhead{}                         &
      \multicolumn{2}{c}{Photometry}     &
      \colhead{}                         &
      \multicolumn{3}{c}{Spectra}        \\ \cline{1-2}\cline{4-6}\cline{8-9}\cline{11-13}
      \colhead{JD}                       &
      \colhead{mag}                      &
      \colhead{}                         &
      \colhead{JD}                       &
      \colhead{$F_{5100}$}               &
      \colhead{$F_{\rm H\beta}$}         &
      \colhead{}                         &
      \colhead{JD}                       &
      \colhead{mag}                      &
      \colhead{}                         &
      \colhead{JD}                       &
      \colhead{$F_{5100}$}               &
      \colhead{$F_{\rm H\beta}$}
         }
\startdata
20.379 & $15.287\pm 0.002$  & &   22.294 & $18.134\pm 0.199$ & $12.601\pm 0.111$ & &   22.327 & $16.631\pm 0.009$  & &   35.306 & $ 4.340\pm 0.031$ & $ 3.638\pm 0.041$ \\
22.273 & $15.297\pm 0.004$  & &   34.307 & $18.091\pm 0.093$ & $12.697\pm 0.089$ & &   39.452 & $16.655\pm 0.011$  & &   40.308 & $ 4.245\pm 0.032$ & $ 3.591\pm 0.047$ \\
39.335 & $15.264\pm 0.002$  & &   39.365 & $18.210\pm 0.047$ & $12.685\pm 0.084$ & &   40.275 & $16.663\pm 0.004$  & &   46.294 & $ 4.114\pm 0.022$ & $ 3.539\pm 0.035$ \\
42.259 & $15.259\pm 0.002$  & &   42.278 & $18.192\pm 0.046$ & $12.688\pm 0.085$ & &   46.260 & $16.664\pm 0.005$  & &   50.345 & $ 4.226\pm 0.031$ & $ 3.565\pm 0.055$ \\
47.215 & $15.266\pm 0.002$  & &   47.239 & $17.938\pm 0.060$ & $12.748\pm 0.077$ & &   50.321 & $16.684\pm 0.006$  & &   68.430 & $ 4.082\pm 0.034$ & $ 3.480\pm 0.066$ 
\enddata
  \tablecomments{\footnotesize
      JD: Julian dates from 2,457,300; $F_{5100}$ and $F_{\hb}$ are the continuum fluxes at 5100 \AA\ and
      \hb\ fluxes in units of $10^{-16}\ {\rm erg\ s^{-1}\ cm^{-2}\ \AA^{-1}}$ and
      $10^{-14}\ {\rm erg\ s^{-1}\ cm^{-2}}$, respectively. This table is available in its entirety in a machine-readable form in the online journal. A portion is shown here for guidance regarding its form and content.
      }
\end{deluxetable*}
\end{table*}

\begin{table*}
\begin{deluxetable*}{rlcrllcrlcrll}
  \tablecolumns{13}
  \setlength{\tabcolsep}{5pt}
  \tablewidth{0pc}
  \tablecaption{Light curves of SDSS~J075101 and SDSS~J075949\label{tab:075101_075949}}
  \tabletypesize{\scriptsize}
  \tablehead{
      \multicolumn{6}{c}{SDSS~J075101}        &
      \colhead{}                         &
      \multicolumn{6}{c}{SDSS~J075949}        \\ \cline{1-6}\cline{8-13}
      \multicolumn{2}{c}{Photometry}     &
      \colhead{}                         &
      \multicolumn{3}{c}{Spectra}        &
      \colhead{}                         &
      \multicolumn{2}{c}{Photometry}     &
      \colhead{}                         &
      \multicolumn{3}{c}{Spectra}        \\ \cline{1-2}\cline{4-6}\cline{8-9}\cline{11-13}
      \colhead{JD}                       &
      \colhead{mag}                      &
      \colhead{}                         &
      \colhead{JD}                       &
      \colhead{$F_{5100}$}               &
      \colhead{$F_{\rm H\beta}$}         &
      \colhead{}                         &
      \colhead{JD}                       &
      \colhead{mag}                      &
      \colhead{}                         &
      \colhead{JD}                       &
      \colhead{$F_{5100}$}               &
      \colhead{$F_{\rm H\beta}$}
         }
\startdata
 372.376 & $16.484\pm 0.003$  & &  383.349 & $ 7.762\pm 0.032$ & $ 4.704\pm 0.037$ & &   32.262 & $17.420\pm 0.011$  & &   32.298 & $ 2.721\pm 0.022$ & $ 2.505\pm 0.044$ \\
 381.415 & $16.440\pm 0.003$  & &  402.305 & $ 7.932\pm 0.020$ & $ 4.751\pm 0.027$ & &   40.338 & $17.440\pm 0.005$  & &   36.308 & $ 2.720\pm 0.023$ & $ 2.371\pm 0.049$ \\
 383.314 & $16.430\pm 0.004$  & &  405.295 & $ 8.144\pm 0.119$ & $ 4.762\pm 0.042$ & &   43.259 & $17.455\pm 0.006$  & &   40.372 & $ 2.560\pm 0.032$ & $ 2.368\pm 0.035$ \\
 402.274 & $16.421\pm 0.002$  & &  408.297 & $ 8.210\pm 0.060$ & $ 4.493\pm 0.058$ & &   48.260 & $17.467\pm 0.006$  & &   48.295 & $ 2.634\pm 0.024$ & $ 2.354\pm 0.028$ \\
 405.267 & $16.419\pm 0.004$  & &  413.436 & $ 8.402\pm 0.025$ & $ 4.854\pm 0.031$ & &   52.316 & $17.474\pm 0.016$  & &   52.343 & $ 2.562\pm 0.043$ & $ 2.274\pm 0.059$ 
\enddata
  \tablecomments{\footnotesize
      JD: Julian dates from 2,457,300; $F_{5100}$ and $F_{\hb}$ are the continuum fluxes at 5100 \AA\ and
      \hb\ fluxes in units of $10^{-16}\ {\rm erg\ s^{-1}\ cm^{-2}\ \AA^{-1}}$ and
      $10^{-14}\ {\rm erg\ s^{-1}\ cm^{-2}}$, respectively. This table is available in its entirety in a machine-readable form in the online journal. A portion is shown here for guidance regarding its form and content.
      }
\end{deluxetable*}
\end{table*}

\begin{table*}
\begin{deluxetable*}{rlcrllcrlcrll}
  \tablecolumns{13}
  \setlength{\tabcolsep}{5pt}
  \tablewidth{0pc}
  \tablecaption{Light curves of SDSS~J081441 and SDSS~J083553\label{tab:081441_083553}}
  \tabletypesize{\scriptsize}
  \tablehead{
      \multicolumn{6}{c}{SDSS~J081441}        &
      \colhead{}                         &
      \multicolumn{6}{c}{SDSS~J083553}        \\ \cline{1-6}\cline{8-13}
      \multicolumn{2}{c}{Photometry}     &
      \colhead{}                         &
      \multicolumn{3}{c}{Spectra}        &
      \colhead{}                         &
      \multicolumn{2}{c}{Photometry}     &
      \colhead{}                         &
      \multicolumn{3}{c}{Spectra}        \\ \cline{1-2}\cline{4-6}\cline{8-9}\cline{11-13}
      \colhead{JD}                       &
      \colhead{mag}                      &
      \colhead{}                         &
      \colhead{JD}                       &
      \colhead{$F_{5100}$}               &
      \colhead{$F_{\rm H\beta}$}         &
      \colhead{}                         &
      \colhead{JD}                       &
      \colhead{mag}                      &
      \colhead{}                         &
      \colhead{JD}                       &
      \colhead{$F_{5100}$}               &
      \colhead{$F_{\rm H\beta}$}
         }
\startdata
 372.380 & $17.538\pm 0.005$  & &  386.419 & $ 2.616\pm 0.021$ & $ 2.362\pm 0.025$ & &   32.409 & $17.225\pm 0.007$  & &   47.340 & $ 3.389\pm 0.016$ & $ 1.870\pm 0.033$ \\
 381.419 & $17.540\pm 0.008$  & &  402.361 & $ 2.712\pm 0.027$ & $ 2.561\pm 0.024$ & &   41.318 & $17.194\pm 0.007$  & &   51.329 & $ 3.567\pm 0.039$ & $ 1.890\pm 0.056$ \\
 383.377 & $17.530\pm 0.010$  & &  406.331 & $ 2.684\pm 0.034$ & $ 2.636\pm 0.045$ & &   47.322 & $17.162\pm 0.004$  & &   67.384 & $ 3.526\pm 0.030$ & $ 1.787\pm 0.072$ \\
 386.391 & $17.470\pm 0.011$  & &  409.303 & $ 2.545\pm 0.027$ & $ 2.504\pm 0.038$ & &   51.300 & $17.171\pm 0.008$  & &   71.340 & $ 3.606\pm 0.045$ & $ 2.074\pm 0.066$ \\
 402.333 & $17.465\pm 0.003$  & &  414.337 & $ 2.699\pm 0.019$ & $ 2.548\pm 0.025$ & &   66.444 & $17.148\pm 0.006$  & &   77.278 & $ 3.581\pm 0.026$ & $ 2.008\pm 0.055$ 
\enddata
  \tablecomments{\footnotesize
      JD: Julian dates from 2,457,300; $F_{5100}$ and $F_{\hb}$ are the continuum fluxes at 5100 \AA\ and
      \hb\ fluxes in units of $10^{-16}\ {\rm erg\ s^{-1}\ cm^{-2}\ \AA^{-1}}$ and
      $10^{-14}\ {\rm erg\ s^{-1}\ cm^{-2}}$, respectively. This table is available in its entirety in a machine-readable form in the online journal. A portion is shown here for guidance regarding its form and content.
      }
\end{deluxetable*}
\end{table*}

\begin{table*}
\begin{deluxetable*}{rlcrllcrlcrll}
  \tablecolumns{13}
  \setlength{\tabcolsep}{5pt}
  \tablewidth{0pc}
  \tablecaption{Light curves of SDSS~J084533 and SDSS~J093302\label{tab:084533_093302}}
  \tabletypesize{\scriptsize}
  \tablehead{
      \multicolumn{6}{c}{SDSS~J084533}        &
      \colhead{}                         &
      \multicolumn{6}{c}{SDSS~J093302}        \\ \cline{1-6}\cline{8-13}
      \multicolumn{2}{c}{Photometry}     &
      \colhead{}                         &
      \multicolumn{3}{c}{Spectra}        &
      \colhead{}                         &
      \multicolumn{2}{c}{Photometry}     &
      \colhead{}                         &
      \multicolumn{3}{c}{Spectra}        \\ \cline{1-2}\cline{4-6}\cline{8-9}\cline{11-13}
      \colhead{JD}                       &
      \colhead{mag}                      &
      \colhead{}                         &
      \colhead{JD}                       &
      \colhead{$F_{5100}$}               &
      \colhead{$F_{\rm H\beta}$}         &
      \colhead{}                         &
      \colhead{JD}                       &
      \colhead{mag}                      &
      \colhead{}                         &
      \colhead{JD}                       &
      \colhead{$F_{5100}$}               &
      \colhead{$F_{\rm H\beta}$}
         }
\startdata
384.333 & $17.692\pm 0.004$  & &  382.430 & $ 1.712\pm 0.030$ & $ 1.374\pm 0.038$ & &  381.425 & $16.920\pm 0.004$  & &  383.413 & $ 4.830\pm 0.026$ & $ 1.401\pm 0.031$ \\
403.406 & $17.721\pm 0.003$  & &  384.365 & $ 1.757\pm 0.012$ & $ 1.246\pm 0.021$ & &  383.384 & $16.931\pm 0.005$  & &  385.389 & $ 4.572\pm 0.029$ & $ 1.138\pm 0.037$ \\
405.430 & $17.718\pm 0.003$  & &  403.438 & $ 1.734\pm 0.016$ & $ 1.323\pm 0.023$ & &  385.377 & $16.947\pm 0.004$  & &  397.442 & $ 4.688\pm 0.051$ & $ 1.245\pm 0.065$ \\
410.298 & $17.761\pm 0.006$  & &  410.329 & $ 1.695\pm 0.014$ & $ 1.228\pm 0.027$ & &  397.413 & $16.955\pm 0.003$  & &  402.425 & $ 4.716\pm 0.019$ & $ 1.324\pm 0.020$ \\
412.309 & $17.767\pm 0.005$  & &  412.348 & $ 1.653\pm 0.012$ & $ 1.263\pm 0.022$ & &  402.403 & $16.950\pm 0.004$  & &  404.389 & $ 4.399\pm 0.017$ & $ 1.254\pm 0.021$
\enddata
  \tablecomments{\footnotesize
      JD: Julian dates from 2,457,300; $F_{5100}$ and $F_{\hb}$ are the continuum fluxes at 5100 \AA\ and
      \hb\ fluxes in units of $10^{-16}\ {\rm erg\ s^{-1}\ cm^{-2}\ \AA^{-1}}$ and
      $10^{-14}\ {\rm erg\ s^{-1}\ cm^{-2}}$, respectively. This table is available in its entirety in a machine-readable form in the online journal. A portion is shown here for guidance regarding its form and content.
      }
\end{deluxetable*}
\end{table*}

\begin{table*}
\begin{deluxetable*}{rlcrllcrlcrll}
  \tablecolumns{13}
  \setlength{\tabcolsep}{5pt}
  \tablewidth{0pc}
  \tablecaption{Light curves of SDSS~J100402 and SDSS~J101000\label{tab:100402_101000}}
  \tabletypesize{\scriptsize}
  \tablehead{
      \multicolumn{6}{c}{SDSS~J100402}        &
      \colhead{}                         &
      \multicolumn{6}{c}{SDSS~J101000}        \\ \cline{1-6}\cline{8-13}
      \multicolumn{2}{c}{Photometry}     &
      \colhead{}                         &
      \multicolumn{3}{c}{Spectra}        &
      \colhead{}                         &
      \multicolumn{2}{c}{Photometry}     &
      \colhead{}                         &
      \multicolumn{3}{c}{Spectra}        \\ \cline{1-2}\cline{4-6}\cline{8-9}\cline{11-13}
      \colhead{JD}                       &
      \colhead{mag}                      &
      \colhead{}                         &
      \colhead{JD}                       &
      \colhead{$F_{5100}$}               &
      \colhead{$F_{\rm H\beta}$}         &
      \colhead{}                         &
      \colhead{JD}                       &
      \colhead{mag}                      &
      \colhead{}                         &
      \colhead{JD}                       &
      \colhead{$F_{5100}$}               &
      \colhead{$F_{\rm H\beta}$}
         }
\startdata
33.422 & $15.574\pm 0.005$  & &   34.421 & $11.915\pm 0.030$ & $ 8.139\pm 0.074$ & &   35.416 & $16.851\pm 0.004$  & &   42.430 & $ 4.283\pm 0.069$ & $ 2.206\pm 0.067$ \\
34.398 & $15.576\pm 0.005$  & &   39.411 & $11.558\pm 0.050$ & $ 8.230\pm 0.084$ & &   42.414 & $16.804\pm 0.005$  & &   48.400 & $ 4.357\pm 0.035$ & $ 2.422\pm 0.031$ \\
39.383 & $15.583\pm 0.004$  & &   43.351 & $11.387\pm 0.073$ & $ 7.893\pm 0.093$ & &   48.369 & $16.807\pm 0.003$  & &   56.349 & $ 4.381\pm 0.068$ & $ 2.403\pm 0.084$ \\
43.330 & $15.600\pm 0.005$  & &   49.392 & $11.684\pm 0.041$ & $ 8.223\pm 0.054$ & &   52.374 & $16.802\pm 0.010$  & &   67.324 & $ 4.621\pm 0.034$ & $ 2.276\pm 0.060$ \\
49.365 & $15.600\pm 0.003$  & &   51.431 & $11.630\pm 0.076$ & $ 8.189\pm 0.066$ & &   67.251 & $16.794\pm 0.008$  & &   78.342 & $ 4.528\pm 0.064$ & $ 2.420\pm 0.042$ 
\enddata
  \tablecomments{\footnotesize
      JD: Julian dates from 2,457,300; $F_{5100}$ and $F_{\hb}$ are the continuum fluxes at 5100 \AA\ and
      \hb\ fluxes in units of $10^{-16}\ {\rm erg\ s^{-1}\ cm^{-2}\ \AA^{-1}}$ and
      $10^{-14}\ {\rm erg\ s^{-1}\ cm^{-2}}$, respectively. This table is available in its entirety in a machine-readable form in the online journal. A portion is shown here for guidance regarding its form and content.
      }
\end{deluxetable*}
\end{table*}

\subsection{Cross-correlation Function}
\label{sec:CCF}

The interpolated cross-correlation function (ICCF; \citealt{gaskell1986,
gaskell1987}) is adopted to determine the time delays of the \hb\ emission
lines to the variation of the continuum light curves. We use the centroid of
the part higher than a threshold (80\% used here) in CCF as the measurement
($\tau_{\rm cent}$) of \hb\ time lag. The uncertainties of the lags are
provided by the ``flux randomization/random subset sampling (FR/RSS)'' method,
which takes into account both the errors of the points in the light curves and
the uncertainties caused by the sampling/cadence for each individual object
\cite[see more details in][]{peterson1998, peterson2004}.  We adopt
$\tau_{\rm cent}$ from the CCFs themselves instead of the median/mean values of
the cross-correlation centroid distributions (CCCDs) produced by the FR/RSS method,
to avoid any undiscovered bias to $\tau_{\rm cent}$ introduced by this method (occasionally, it 
overestimates the uncertainties, e.g., \citealt{peterson1998}). The CCCDs are
only used to estimate the error bars. But it should be noted that, in the
present sample, the $\tau_{\rm cent}$ from the
CCFs and the median/mean values of the CCCDs are highly consistent. Their differences are significantly smaller than
the error bars (typically $\lesssim10\%$ of the error bars), and can be ignored. 

The auto-correlation functions (ACFs), CCFs
and the distributions of the centroid time lags in the observed frame are shown in Figure
\ref{fig:light_curves}. The time lags, their uncertainties, and the
corresponding maximum correlation coefficients ($r_{\rm max}$) are listed in
Table \ref{tab:lags}. For the objects monitored for more than one year, we
also measured the \hb\ lags from the light curves only in 2015 Oct - 2016 Jun
or in 2016 Oct - 2017 Jun. In some cases the $r_{\rm max}$ measured from the
light curves in a single year is significantly higher than the values
calculated from all the light curves,  or uncertainties of the \hb\ lags
are smaller because their ACFs are much narrower. We tend to use the time lags
determined from the data in a single year,  because the gaps in the light
curves between the two years may introduce some uncertainties in the lag
measurements. We discuss the light curves and the time delays for 
individual objects in the Section \ref{sec:individual_obj}. The \hb\ time lags we
used in the measurements of their BH masses are labeled by ``$\surd$'' in
Table \ref{tab:lags}.

\begin{table}
\begin{deluxetable}{lcclclc}
  \tablecolumns{6}
  \setlength{\tabcolsep}{4pt}
  \tablewidth{0pc}
  \tablecaption{Time Lags\label{tab:lags}}
  \tabletypesize{\scriptsize}
  \tablehead{
      \colhead{}                         &
      \colhead{}                         &
      \colhead{}                         &
      \colhead{Observed}                   &
      \colhead{}                         &
      \colhead{Rest-frame}                   &
      \colhead{}           \\
      \colhead{Object}                   &
      \colhead{Period}                   &
      \colhead{$r_{\rm max}$}            &
      \colhead{Time Lag}                 &
      \colhead{}                         &
      \colhead{Time Lag}                 &
      \colhead{Note}                       \\ 
      \colhead{}                         &
      \colhead{}                         &
      \colhead{}                         &
      \colhead{(days)}                   &
      \colhead{}                         &
      \colhead{(days)}                   &
      \colhead{}                         
            }
\startdata
SDSS~J074352 & 2015-2017 & 0.94 & $68.6_{-10.2}^{+ 4.7}$ & & $54.8_{- 8.1}^{+ 3.7}$ &         \\
            & 2016-2017 & 0.79 & $55.0_{- 5.2}^{+ 6.6}$ & & $43.9_{- 4.2}^{+ 5.2}$ & $\surd$ \\
SDSS~J075051 & 2015-2017 & 0.69 & $93.3_{-13.9}^{+26.2}$ & & $66.6_{- 9.9}^{+18.7}$ & $\surd$ \\
SDSS~J075101 & 2016-2017 & 0.96 & $32.0_{- 7.6}^{+ 6.3}$ & & $28.6_{- 6.8}^{+ 5.6}$ & $\surd$ \\
SDSS~J075949 & 2014-2017 & 0.74 & $47.0_{-12.1}^{+10.3}$ & & $39.5_{-10.2}^{+ 8.7}$ &         \\
            & 2015-2016 & 0.83 & $31.3_{-11.3}^{+13.8}$ & & $26.4_{- 9.5}^{+11.6}$ & $\surd$ \\
SDSS~J081441 & 2016-2017 & 0.76 & $31.2_{- 6.9}^{+ 8.4}$ & & $26.8_{- 5.9}^{+ 7.3}$ & $\surd$ \\
SDSS~J083553 & 2015-2017 & 0.85 & $42.6_{- 7.6}^{+ 7.1}$ & & $35.4_{- 6.3}^{+ 5.9}$ &         \\
            & 2016-2017 & 0.86 & $14.9_{- 6.6}^{+ 6.5}$ & & $12.4_{- 5.4}^{+ 5.4}$ & $\surd$ \\
SDSS~J084533 & 2016-2017 & 0.78 & $25.9_{- 5.1}^{+ 9.5}$ & & $19.9_{- 3.9}^{+ 7.3}$ & $\surd$ \\
SDSS~J093302 & 2016-2017 & 0.66 & $22.4_{- 5.0}^{+ 4.5}$ & & $19.0_{- 4.3}^{+ 3.8}$ & $\surd$ \\
SDSS~J100402 & 2015-2017 & 0.84 & $ 1.4_{-19.0}^{+31.6}$ & & $ 1.0_{-14.3}^{+23.8}$ &         \\
            & 2016-2017 & 0.60 & $42.8_{- 5.5}^{+57.7}$ & & $32.2_{- 4.2}^{+43.5}$ & $\surd$ \\
SDSS~J101000 & 2015-2017 & 0.81 & $48.1_{-10.5}^{+19.0}$ & & $38.2_{- 8.4}^{+15.1}$ &         \\
            & 2016-2017 & 0.61 & $34.9_{- 9.6}^{+29.5}$ & & $27.7_{- 7.6}^{+23.5}$ & $\surd$ 
 \enddata
  \tablecomments{\footnotesize
  ``$\surd$'' means we use this time lag of the object to calculate its BH mass.
      The lag of SDSS~J075051 is obtained from its photometric and \hb\ light curves. }
\end{deluxetable}
\end{table}

\subsection{Notes on Individual Objects}
\label{sec:individual_obj}

{\it SDSS~J074352:} Both the continuum and \hb\ light curves show very clear
dips during 2016--2017. In general, the time lag measured from the
light curves in 2016--2017 is consistent with the lag measured from its entire
light curves, within the uncertainties. However, the ACF generated solely from its
continuum light curve in 2016--2017 is much narrower than the ACF obtained
from its entire continuum light curve. 
Considering that the season gap between
2016 Jun to 2016 Oct may influence the \hb\ lag measurement, we adopt
the CCF analysis of the light curves in 2016--2017 as the final result
of this object.

{\it SDSS~J075051:} The quality of its photometric light curve is 
superior to the 5100 \AA\ continuum light curve.  The $r_{\rm max}$ of
photometry vs. \hb\ is higher than the value ($\sim 0.4$) of
5100 \AA\ vs. \hb. More observations are needed to improve its lag measurement in
the future. 

{\it SDSS~J075101:} It was monitored previously in 2013 Nov -- 2014 May (Paper
\citetalias{du2015}), yielding an \hb\ time lag of $33.4_{-5.6}^{+15.6}$ days in the rest frame.
The new measurement in the present paper is consistent with the previous
observations, within the uncertainties, but the error bars are smaller. The peak
around Julian date 550 (from the zero point of 2457300 in Figure \ref{fig:light_curves}, similarly hereinafter) 
is prominent, and its $r_{\rm max}$ is very high
(0.96).

{\it SDSS~J075949:} We have monitored it for three years, from 2014 to 2017.
The result of the first year was published in Paper \citetalias{du2016V}, 
showing relatively large error bars in its \hb\ time delay
($55.0_{-13.1}^{+17.0}$ days in the rest frame). We also plot its old light curves in 2014--2015 in Figure \ref{fig:light_curves}
for comparison. 
The new observation in this paper gives better
constraints on the \hb\ lag.  The light curves in 2015--2016 give  a higher
correlation coefficient ($r_{\rm max}=0.83$) than the value ($r_{\rm max}=0.74$)
obtained from its entire light curves (including the data in 2014--2015).  We
thus use the CCF results from 2015--2016 in the analysis of the \rl\ relation.
The new time lag is $26.4_{-9.5}^{+11.6}$ days in the rest frame. This is 
somewhat shorter than the previous value (Paper \citetalias{du2016V}), but considering the uncertainties, the difference is not significant.

{\it SDSS~J081441:} Its light curves from 2013 Nov -- 2014 May were published in 
Paper \citetalias{du2015}. In the \hb\ 2013 -- 2014 light curve, there are 
only several points after the peak around Julian day 160 (see Figure 1 of Paper \citetalias{du2015}),
 because the altitude 
of the source was already too low to observe at the end. The new 2016 -- 2017 light curves look
more convincing, and the lag measurement is much better (with smaller error bars). 

{\it SDSS~J083553:} It is a little unfortunate that the primary peaks around 
Julian day 
$\sim$300 in the continuum and \hb\ light curves are invisible.
However, the small dip close to Julian day $\sim$450 is observed clearly.
So, we adopt the CCF analysis from the 2016--2017 data to be the final result for this object. 
Its ACF is narrower and the $r_{\rm max}$ is a little higher than the analysis obtained
from the entire light curves from 2015 to 2017. 

{\it SDSS~J084533:} The time lag in the present paper is consistent  with
the previous value from 2014 -- 2015 (Paper
\citetalias{du2016V}). Its light curves from 2016 to 2017 show two large
structures, a dip around Julian day 430 and a peak around Julian day
480, yielding a very robust lag measurement. 

{\it SDSS~J093302:} The big dip and its response are very clear in 
the light curves. The lag measurement is pretty reliable.

{\it SDSS~J100402:} The centroid of the peak of the CCF calculated from the entire light curves is nearly zero because of the the gap between the two years and the very flat light curves in 2016 -- 2017. 
Therefore, we adopt the light curves
in 2015 -- 2016 to measure the time lag. The uncertainty of its \hb\ lag is the 
largest among all of the objects in this paper. 

{\it SDSS~J101000:} The scatter and the error bars of the \hb\
light curve in 2016 -- 2017 are smaller than those in 2015 -- 2016. Thus, we 
select the lag measurement using the light curves in 2016 -- 2017 as the final result. 
In general, the lags obtained from the light curves in 2015 -- 2017 and in 2016 -- 2017
are consistent with each other.

\begin{table}
\begin{deluxetable}{lccccc}
  \tablecolumns{6}
  \setlength{\tabcolsep}{3pt}
  \tablewidth{0pc}
  \tablecaption{\hb\ Width Measurements\label{tab:fwhm}}
  \tabletypesize{\scriptsize}
  \tablehead{
      \colhead{}                   &
      \multicolumn{2}{c}{Mean Spectra}   &
      \colhead{}                         &
      \multicolumn{2}{c}{RMS Spectra}        \\ \cline{2-3} \cline{5-6}
      \colhead{Object}                         &
      \colhead{FWHM}                     &
      \colhead{$\sigma_{\hb}$}           &
      \colhead{}                         &
      \colhead{FWHM}                     &
      \colhead{$\sigma_{\hb}$}           \\
      \colhead{}                         &
      \colhead{(km s$^{-1}$)}            &
      \colhead{(km s$^{-1}$)}            &
      \colhead{}                         &
      \colhead{(km s$^{-1}$)}            &
      \colhead{(km s$^{-1}$)}                }
\startdata
SDSS~J074352 & $3156\pm36$ & $1976\pm10$ & & $3149\pm  92$ & $1489\pm 32$ \\
SDSS~J075051 & $1904\pm 9$ & $1239\pm 5$ & & $ 970\pm 345$ & $ 547\pm 46$ \\
SDSS~J075101 & $1679\pm35$ & $1179\pm14$ & & $1605\pm 630$ & $ 987\pm 74$ \\
SDSS~J075949 & $1783\pm17$ & $1135\pm 3$ & & $1661\pm 402$ & $ 845\pm 40$ \\
SDSS~J081441 & $1782\pm16$ & $1367\pm 6$ & & $1247\pm1048$ & $1195\pm379$ \\
SDSS~J083553 & $1758\pm16$ & $1015\pm 9$ & & $1642\pm 479$ & $1234\pm 15$ \\
SDSS~J084533 & $1297\pm12$ & $ 965\pm 7$ & & $1626\pm  98$ & $1126\pm  9$ \\
SDSS~J093302 & $1800\pm25$ & $1423\pm 7$ & & $1526\pm 313$ & $ 838\pm283$ \\
SDSS~J100402 & $2088\pm 1$ & $1425\pm 5$ & & $2555\pm  78$ & $1173\pm132$ \\
SDSS~J101000 & $2311\pm 1$ & $1409\pm 1$ & & $2237\pm 129$ & $1386\pm150$
 \enddata
\end{deluxetable}
\end{table}

\subsection{Contribution of Host Galaxies} 
\label{sec:host}

Generally, the contribution of host galaxies in the slit can be decomposed and
removed from the 5100 \AA\ luminosities of the objects by using the high-resolution
image observations (e.g., from Hubble Space Telescope, {\it HST}).
However, none of the objects, except SDSS~J100402 (also known as PG 1001+291),
has imaging observations from {\it HST}. As in Papers \citetalias{du2015} and
\citetalias{du2016V},  we uniformly adopt the empirical relation proposed by
\cite{shen2011}, to remove the host contribution in the 5100 \AA\ luminosities.
The luminosity ratio of host to AGN at 5100 \AA\ can be expressed as
$L_{5100}^{\rm host}/L_{5100}^{\rm AGN}=0.8052-1.5502x+0.912x^2-0.1577x^3$,
for $x<1.053$,  where $x=\log \left(L_{5100}^{\rm tot}/10^{44}{\rm
erg~s^{-1}}\right)$ and  $L_{5100}^{\rm tot}$ is the total luminosity at 5100
 \AA. For $x>1.053$,  $L_{5100}^{\rm host}\ll L_{5100}^{\rm AGN}$, and the host
contribution can be ignored. The fractions of host contamination in the total 5100 \AA\
luminosities are 26.0\%, 28.0\%, 37.2\%, 17.9\%, 14.9\%, 23.1\%, and
6.8\% for SDSS~J075101, J075949, J081441, J083553, J084533, J093302,
and J101000, respectively. The host contribution can be ignored for
SDSS~J074352, J075051, and J100402.

\subsection{Black Hole Masses and Accretion Rates}
\label{sec:mass_mdot}
 
We use Equation (\ref{eqn:mbh}) to calculate the BH masses of the objects
observed in SEAMBH2015--2016.  The widths of the \hb\ emission lines can be
obtained from the FWHM or $\sigma_{\hb}$ measured from their mean or RMS spectra.  Different works adopt
different line width measurements (e.g., \citealt{peterson2004, bentz2009,
denney2010, grier2012, kaspi2005}, Papers
\citetalias{du2014}-\citetalias{du2016V}). In general, the BH masses produced
by the different line width measurements are consistent, because their
corresponding virial factors \fblr\ are calibrated, in the same way, by
comparing the RM objects with measurements of bulge stellar velocity
dispersion ($\sigma_{*}$) with the $\mbh-\sigma_{*}$ relation of inactive
galaxies (e.g., \citealt{onken2004, woo2010, park2012, grier2013, ho2014,
woo2015}, see a brief review in \citealt{du2017}). 
The exact value of \fblr\ is still a matter of some debate and has large uncertainties. 
Recently, \cite{woo2015} found that narrow-line Seyfert 1 (NLS1) galaxies (the \hb\ widths of most SEAMBHs conform to 
the definition of NLS1s) has a value of \fblr\ = 
1.12.  On the other hand, \cite{ho2014} show that \fblr\ is smaller than 1 for the AGNs 
with pseudobulges (NLS1s tend to host pseudobulges; e.g., \citealt{mathur2012}). 
It is not clear will be the final value of \fblr. More observations are needed to calibrate \fblr\ in the future.
At this stage,  as in the other papers in our series
(Papers \citetalias{du2014}-\citetalias{du2016V}), we adopt FWHM measured
from the mean spectra and $\fblr=1$ to calculate the
BH masses, but we acknowledge the large uncertainty on \fblr.

In order to measure the FWHM of broad \hb, the narrow component of the line
should be removed. This is done by fixing the flux of narrow \hb\
to 10\% of the flux of \oiii\ $\lambda5007$ (the typical value in AGNs; e.g., \citealt{stern2013, kewley2006}),  and the uncertainty is
estimated by setting \hb$/$\oiii$\lambda5007$ to 0\% and 20\% as the lower  and
upper limit, respectively. Narrow \hb\ and \oiii\ are weak in SEAMBHs; thus,
the influence of the narrow-component subtraction to the measurements is
not very significant. We measure FWHM of \hb\ from the mean spectra after removing
the narrow component. For completeness, we also provide $\sigma_{\hb}$
measured from the mean spectra, and FWHM and $\sigma_{\hb}$ measured
from the RMS spectra. The widths are measured from the RMS spectra after smoothing by 
a 9-pixel boxcar, and the uncertainties are obtained by comparing with the 
measurements from the profiles smoothed by a 3-pixel boxcar. 
The instrumental broadening (FWHM$\approx$1200 km s$^{-1}$), estimated from the spectra of 
the comparison stars, has been subtracted from the width measurements. 
The \hb\ line width and their uncertainties are listed in Table \ref{tab:fwhm}.

As in Papers \citetalias{du2015} and \citetalias{du2016V}, the dimensionless
accretion rates of the objects are estimated by Equation (\ref{eqn:mdot})
because we cannot observe their entire spectral energy distributions (SEDs).
Equation (\ref{eqn:mdot}) is derived from the thin accretion disk model
(\citealt{shakura1973, frank2002}; see more details in Paper
\citetalias{wang2014}), and can be used as a substitute for the traditional
estimate of Eddington ratio (e.g., $L_{\rm bol}/L_{\rm Edd}=10
L_{5100}/L_{\rm Edd}$, where $L_{\rm bol}$ is the bolometric luminosity and
$L_{\rm Edd}$ is the Eddington Luminosity). Its validity has been discussed in
the Appendix of Paper \citetalias{du2016V}, and it applies to
$\dotm\lesssim3\times10^3 m_7^{-1/2}$, where $m_7=\mbh/10^7M_{\odot}$. We
list the BH masses, 5100 \AA\ and \hb\ luminosities, and the
accretion rates in Table \ref{tab:mbh_mdot}.

As described in Papers \citetalias{wang2014} and \citetalias{du2016V},
the criterion of \dotm\ for identifying SEAMBHs still has some
uncertainties \citep{laor1989, beloborodov1998, sadowski2011}. We can use
$\eta\dotm\ge0.1$ as the criterion, because the accretion disk
becomes slim and the radiation efficiency gets reduced \citep{sadowski2011}, 
where $\eta$ is the mass-to-radiation conversion efficiency.
To be conservative, we adopted the lowest $\eta$
(0.038, for retrograde disk with BH spin $a=-1$; see \citealt{Bardeen1972}).  Thus, SEAMBHs
are AGNs with $\dotm\ge2.63$. For simplicity, and as in other papers in this 
series, we use $\dotm=3$ as the criterion to distinguish SEAMBHs from AGNs 
with low accretion rates.

\begin{figure*}
\centering
\includegraphics[width=0.75\textwidth]{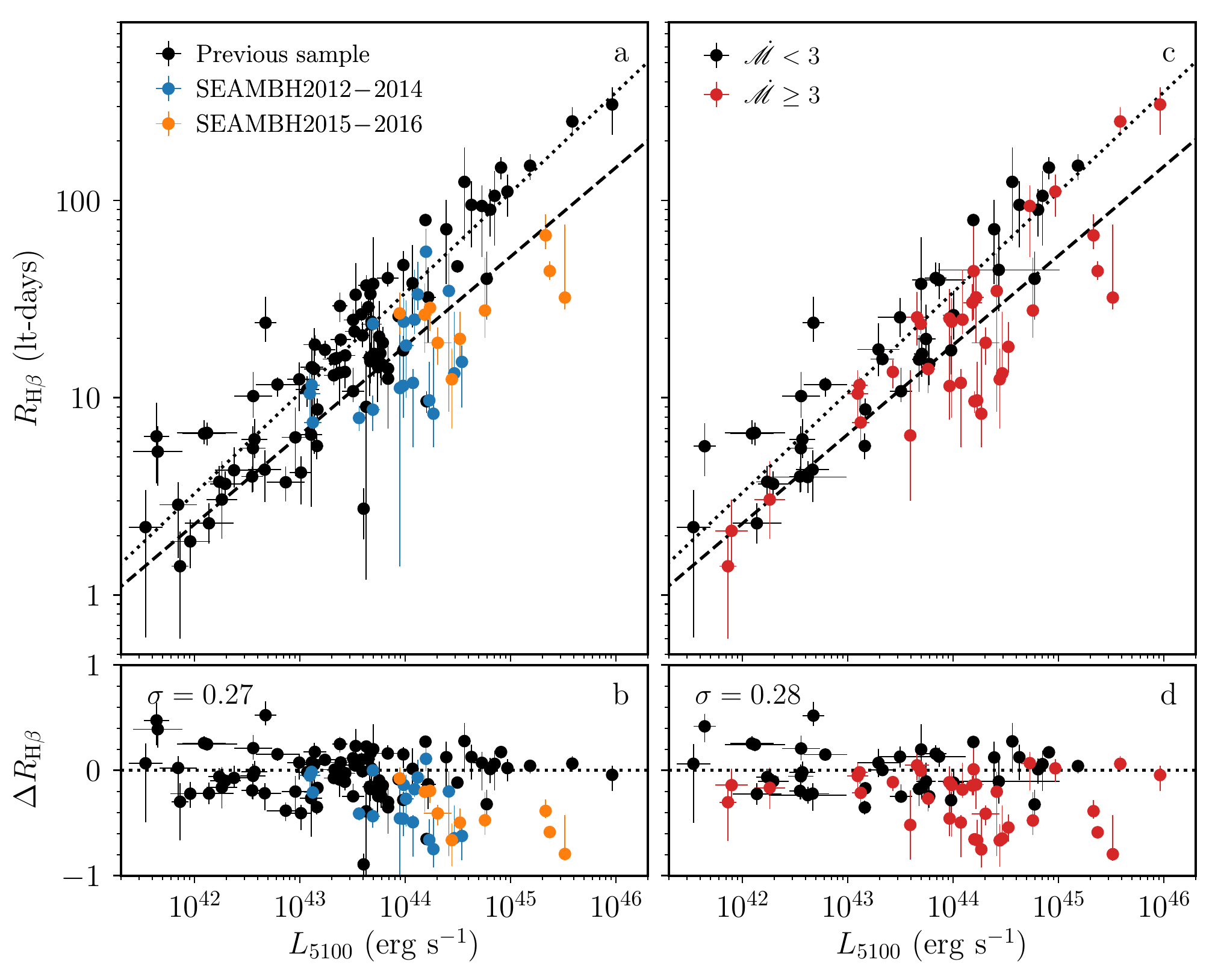} 
\caption{\rl\ relation. In panels a and b, each campaign for the objects
with repeated RM observations is shown as a single point. In panels c and d,
multiple observations for the same object have been averaged.
 The dotted and dashed lines are the linear regressions for
the AGNs with $\dotm<3$ and with $\dotm \ge 3$, respectively. $\sigma$ is the
standard deviation of the residual $\Delta \rhb$. In the left panels, the blue and orange
points are the SEAMBHs observed in our campaign, and the black points are the
other RM objects. In the right panels, the red points are the objects with $\dotm\ge3$, and the 
black ones are the AGNs with low accretion rates.}
\label{fig:r_l}
\end{figure*}

\begin{figure*}
\centering
\includegraphics[width=\textwidth]{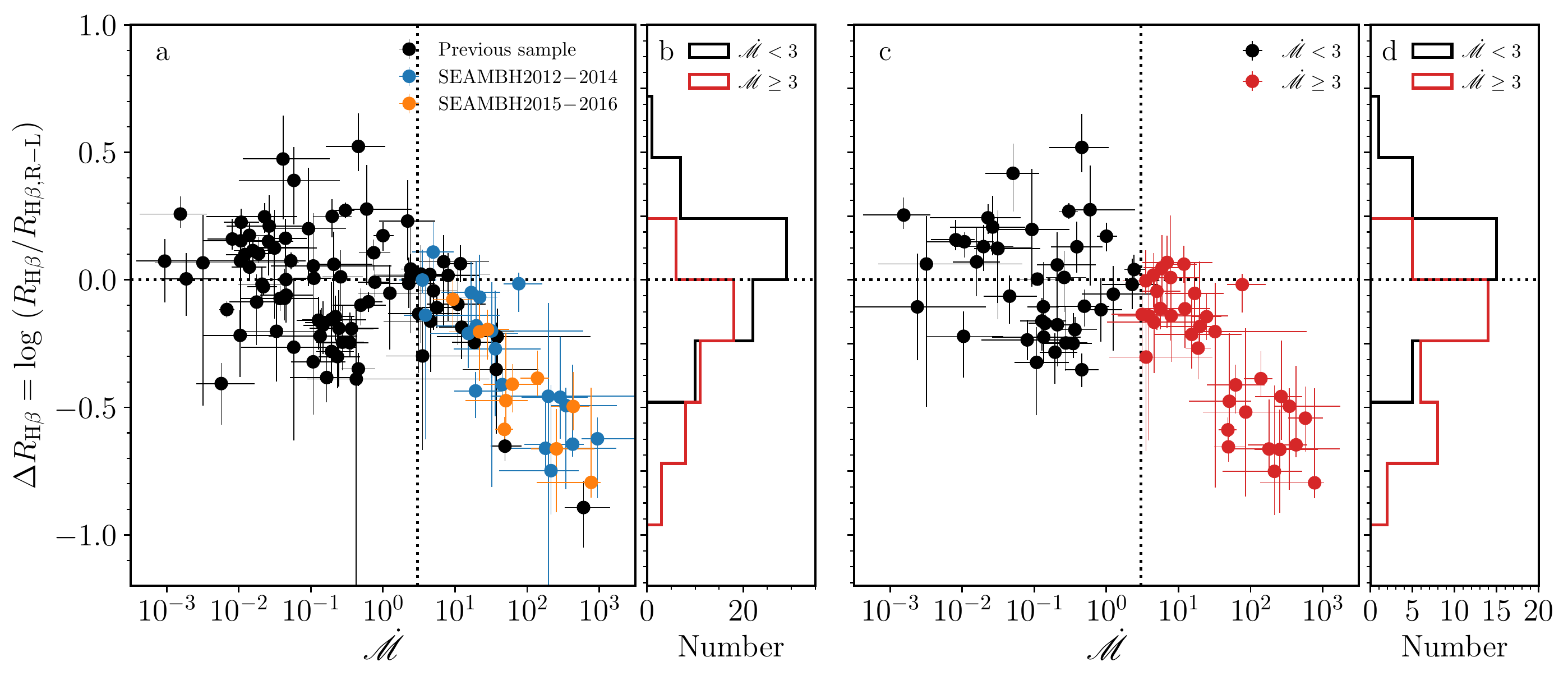} 
\caption{The correlation between $\Delta \rhb$ and \dotm. $\Delta \rhb$ is the
deviations of the SEAMBHs from the \rl\ relation of sub-Eddington sources. The
meaning of the different colors in panels a and c are the same in Figure
\ref{fig:r_l}. Panels b and d show the distributions of $\Delta \rhb$ in
different subsamples in panels a and c. The vertical and horizontal dashed
lines mark $\dotm=3$ and $\Delta \rhb=0$, respectively.}
\label{fig:deltar_mdot}
\end{figure*}

\section{Properties of \hb\ lags in SEAMBHs} 
\label{sec:r-l}

The time delays of the \hb\ emission lines in SEAMBHs have been shown to
be shorter by a factor of 2 -- 8 compared to AGNs with normal
accretion rates, and the degree of shortening strongly correlates with the
accretion rates (Papers \citetalias{du2015} and \citetalias{du2016V}). To
investigate the magnitude of the shortening for the newly observed SEAMBHs
with higher luminosities, we plot the \rl\ relation in Figure \ref{fig:r_l},
including the SEAMBHs newly observed from 2015 to 2017, SEAMBHs published
in Papers \citetalias{du2014} -- \citetalias{du2016V}, other objects
compiled in Papers \citetalias{du2015} and \citetalias{du2016V}, and several
newly observed AGNs studied by other groups\footnote{Besides our new SEAMBH
targets, several new RM objects published after Paper \citetalias{du2016V} are
also included: MCG-06-30-15 from \cite{bentz2016mcg} and
\cite{hu2016}; UGC 06728 from \cite{bentz2016ugc}; MCG+08-11-011, NGC 2617, NGC
4051, 3C 382, and Mrk 374 from \cite{fausnaugh2017}.} published after Paper
\citetalias{du2016V}. In the left panel of Figure \ref{fig:r_l}, each campaign
for the objects with repeated RM observations is treated as a single point
(called ``direct scheme''); in the right panel, multiple observations for 
the same object  (called ``averaged scheme'' see more details in 
Paper \citetalias{du2015}) are averaged.  It is obvious that the sample in the
present paper, in general, is more luminous than that of our previous SEAMBH campaigns
(except for the four objects previously observed in Papers
\citetalias{du2015} and \citetalias{du2016V}). The objects in SEAMBH2015-2016
still show much shorter \hb\ lags than AGNs of the same luminosity with normal accretion rates.

\subsection{The BLR Size-Luminosity Relation} 

Using the FITEXY algorithm as
modified  by \cite{tremaine2002}, we obtain the linear regression for the direct scheme 
\begin{equation}  
\log\frac{\rhb}{{\rm ltd}}\! =\! 
\left\{\!  
             \begin{array}{ll}  
             \!(1.40\pm0.03) \!+\! (0.43\pm0.03)\log \ell_{44}, & \!(\text{all }\dotm) \\  
             \!(1.53\pm0.03) \!+\! (0.51\pm0.03)\log \ell_{44}, & \!(\dotm \!<\! 3)            \\  
             \!(1.26\pm0.04) \!+\! (0.45\pm0.05)\log \ell_{44}, & \!(\dotm \!\ge\! 3)   
             \end{array}  
\right. 
\label{eqn:r-l-direct}
\end{equation}
with intrinsic scatter $\sigma_{\rm in} =$ (0.21, 0.16, 0.22).
For the averaged scheme, we find
\begin{equation}  
\log\frac{\rhb}{{\rm ltd}}\! =\! 
\left\{\!  
             \begin{array}{ll}  
             \!(1.39\pm0.03) \!+\! (0.43\pm0.03)\log \ell_{44}, & \!(\text{all }\dotm) \\  
             \!(1.53\pm0.04) \!+\! (0.51\pm0.04)\log \ell_{44}, & \!(\dotm \!<\! 3)            \\  
             \!(1.27\pm0.05) \!+\! (0.45\pm0.05)\log \ell_{44}, & \!(\dotm \!\ge\! 3)   
             \end{array}  
\right. 
\label{eqn:r-l-averaged}
\end{equation}
with intrinsic scatter $\sigma_{\rm in} =$ (0.22, 0.17, 0.21).
The intercepts of the correlations for the AGNs with high ($\dotm \ge 3$) and
low ($\dotm < 3$) accretion rates are significantly different. On average,
the \hb\ lags of the $\dotm \ge 3$ sources are shorter than the values of the
$\dotm < 3$ sources by a factor of $\sim2$ (0.26 dex). The slope of the 
SEAMBHs may be slightly smaller, although the difference is not very
significant, considering the uncertainties. In view of the currently limited 
and its non-uniform distribution, especially at low and high luminosities, it 
is premature to draw firm conclusions on the slopes of the correlations.  At 
this stage, as a preliminary result, the slopes of high- and low-$\dotm$ 
objects can be regarded as indistinguishable.

\begin{deluxetable*}{lllcccrccl}
 \tablecolumns{8}
 \setlength{\tabcolsep}{10pt}
\tablewidth{\textwidth}
 \tablecaption{Results of H$\beta$ Reverberation Mapping of the SEAMBHs in 2015 -- 2017\label{tab:mbh_mdot}}
 \tabletypesize{\scriptsize}
 \tablehead{
     \colhead{Objects}                  &
     \colhead{$\tau_{\hb}$}                 &
     \colhead{FWHM}                     &
     \colhead{$\log \mbh$}&
     \colhead{$\log \dotm$}         &
     \colhead{$\log L_{5100}$}          &
     \colhead{$\log L_{\hb}$}              &
     \colhead{EW(\hb)}             \\ \cline{2-8}
     \colhead{}                         &
     \colhead{(days)}                   &
     \colhead{(km s$^{-1}$)}                 &
     \colhead{($M_{\odot}$)}                         &
     \colhead{}                         &
     \colhead{(erg s$^{-1}$)}                &
     \colhead{(erg s$^{-1}$)}                &
     \colhead{(\AA)}
 }
\startdata
SDSS~J074352  & $     43.9_{-  4.2}^{+  5.2} $ & $     3156\pm   36 $ & $    7.93_{-0.04}^{+0.05} $ & $      1.69_{-  0.13}^{+  0.12} $ & $    45.37\pm 0.02 $ & $    43.48\pm 0.01 $ & $     65.8\pm  3.5 $  \\ 
SDSS~J075051  & $     66.6_{-  9.9}^{+ 18.7} $ & $     1904\pm    9 $ & $    7.67_{-0.07}^{+0.11} $ & $      2.14_{-  0.24}^{+  0.16} $ & $    45.33\pm 0.01 $ & $    43.34\pm 0.03 $ & $     51.9\pm  4.5 $  \\ 
SDSS~J075101  & $     28.6_{-  6.8}^{+  5.6} $ & $     1679\pm   35 $ & $    7.20_{-0.12}^{+0.08} $ & $      1.45_{-  0.23}^{+  0.30} $ & $    44.24\pm 0.04 $ & $    42.38\pm 0.04 $ & $     70.4\pm  9.0 $  \\ 
SDSS~J075949  & $     26.4_{-  9.5}^{+ 11.6} $ & $     1783\pm   17 $ & $    7.21_{-0.19}^{+0.16} $ & $      1.34_{-  0.42}^{+  0.48} $ & $    44.19\pm 0.06 $ & $    42.47\pm 0.04 $ & $     98.9\pm 17.0 $  \\ 
SDSS~J081441  & $     26.8_{-  5.9}^{+  7.3} $ & $     1782\pm   16 $ & $    7.22_{-0.11}^{+0.10} $ & $      0.97_{-  0.28}^{+  0.28} $ & $    43.95\pm 0.04 $ & $    42.39\pm 0.02 $ & $    140.4\pm 16.2 $  \\ 
SDSS~J083553  & $     12.4_{-  5.4}^{+  5.4} $ & $     1758\pm   16 $ & $    6.87_{-0.25}^{+0.16} $ & $      2.41_{-  0.35}^{+  0.53} $ & $    44.44\pm 0.02 $ & $    42.48\pm 0.02 $ & $     56.1\pm  4.0 $  \\ 
SDSS~J084533  & $     19.9_{-  3.9}^{+  7.3} $ & $     1297\pm   12 $ & $    6.82_{-0.10}^{+0.14} $ & $      2.64_{-  0.31}^{+  0.22} $ & $    44.52\pm 0.02 $ & $    42.60\pm 0.03 $ & $     61.7\pm  5.1 $  \\ 
SDSS~J093302  & $     19.0_{-  4.3}^{+  3.8} $ & $     1800\pm   25 $ & $    7.08_{-0.11}^{+0.08} $ & $      1.79_{-  0.40}^{+  0.40} $ & $    44.31\pm 0.13 $ & $    42.10\pm 0.05 $ & $     31.8\pm 10.3 $  \\ 
SDSS~J100402  & $     32.2_{-  4.2}^{+ 43.5} $ & $     2088\pm    1 $ & $    7.44_{-0.06}^{+0.37} $ & $      2.89_{-  0.75}^{+  0.13} $ & $    45.52\pm 0.01 $ & $    43.54\pm 0.01 $ & $     53.6\pm  1.3 $  \\ 
SDSS~J101000  & $     27.7_{-  7.6}^{+ 23.5} $ & $     2311\pm    1 $ & $    7.46_{-0.14}^{+0.27} $ & $      1.70_{-  0.56}^{+  0.31} $ & $    44.76\pm 0.02 $ & $    42.77\pm 0.02 $ & $     52.6\pm  3.4 $   
\enddata
\tablecomments{\footnotesize 
$\tau_{\hb}$ (in the rest-frame) and FWHM are the same as in Tables \ref{tab:lags} and \ref{tab:fwhm}; we list them here again for the convenience of inspection. $L_{5100}$ are the luminosities corresponding to the light curves used for $\tau_{\hb}$ measurements. The host contribution in $L_{5100}$
has been removed (see Section \ref{sec:host}).
Galactic extinction has been corrected using the maps of \cite{schlafly2011}.}
\end{deluxetable*}

\subsection{Dependence of BLR Size on Accretion Rate}

According to Papers \citetalias{du2015} and \citetalias{du2016V}, the 
shortening of the \hb\ lags in SEAMBHs show a strong correlation with 
accretion rate.  In order
to test whether this correlation extends to SEAMBHs of even higher luminosities, we define, as in Papers \citetalias{du2015} and \citetalias{du2016V},  $\Delta \rhb = \log
(\rhb/R_{{\rm H\beta},R-L})$ to quantify the deviation from the \rl\ relation
of the subsample with $\dotm<3$ (i.e., $R_{{\rm H\beta},R-L}$ is the
correlation for $\dotm<3$ in Equations (\ref{eqn:r-l-direct}) or
(\ref{eqn:r-l-averaged})). Figure \ref{fig:deltar_mdot} shows the correlation
between $\Delta \rhb$ and \dotm, as well as the distributions of the AGNs with
$\dotm \ge 3$ and $\dotm < 3$ in the direct and averaged schemes. The objects
with $\dotm < 3$ are located in both the two left quadrants of $\Delta \rhb
\ge 0$ and $\Delta \rhb < 0$. However, SEAMBHs ($\dotm \ge 3$) only appear in
the quadrant with $\Delta \rhb < 0$, and their $\Delta \rhb$ values significantly correlate
with \dotm\ (Pearson's correlation coefficient and null-probability
are $-0.84$ and $4.8\times10^{-13}$ for the direct scheme; the corresponding 
values are $-0.82$ and $1.4\times10^{-9}$
for the average scheme).  The objects in SEAMBH2015--2016 follow the same
correlation as those SEAMBHs with lower luminosities  observed in our campaign
between 2012 and 2015. The dependence on accretion rate for SEAMBHs of the \rhb\ deviations from the \rl\
relation can be obtained by the regression
for the objects with $\dotm \ge 3$:
\begin{equation}  
\Delta \rhb\! =\! 
\left\{\!  
             \begin{array}{ll}  
             \!(0.36\pm0.08) \!-\! (0.44\pm0.05)\log \dotm, & \!\text{(direct)} \\  
             \!(0.30\pm0.09) \!-\! (0.40\pm0.06)\log \dotm, & \!\text{(averaged)}            
             \end{array}  
\right. 
\label{eqn:deltar_mdot}
\end{equation}
with intrinsic scatter $\sigma_{\rm in} = (0.04, 0.07)$.

\begin{figure*}
\centering
\includegraphics[width=\textwidth]{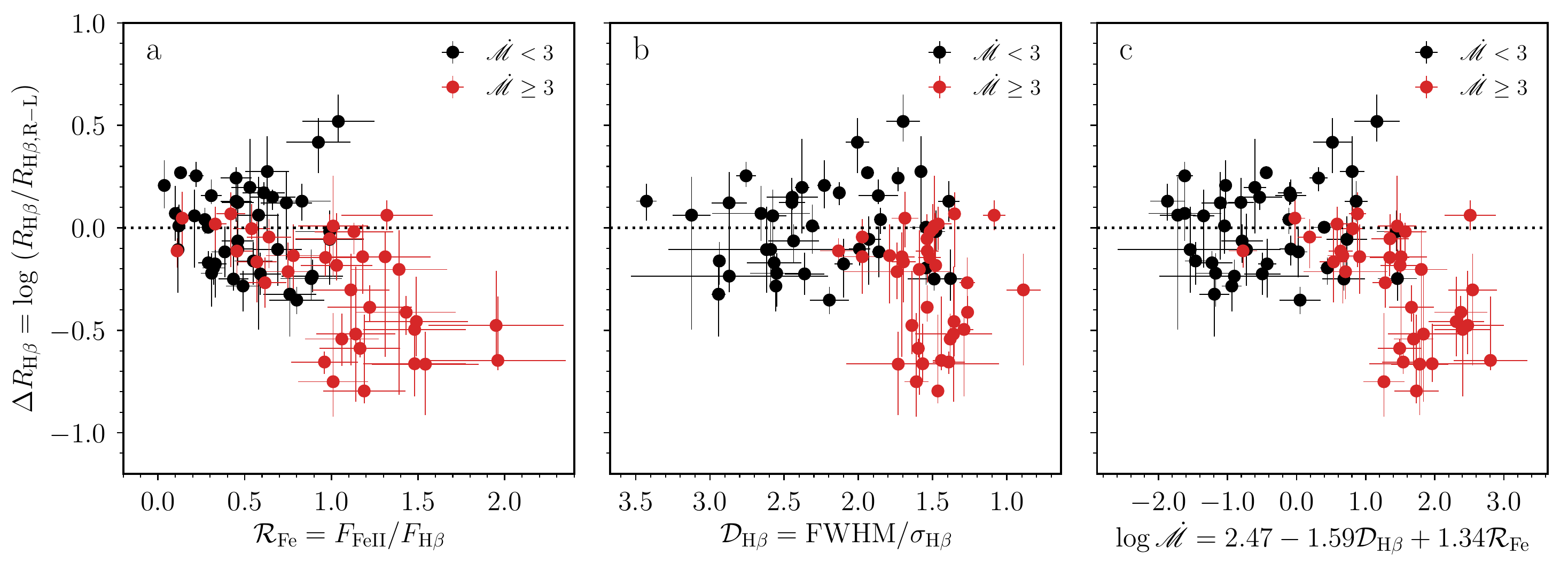} 
\caption{$\Delta \rhb$ vs. \Rfe, \Dhb, and \dotm\ derived from the fundamental
plane. The repeated observations for the same objects have been averaged as in
the right panels of Figures \ref{fig:r_l} and \ref{fig:deltar_mdot} (average
scheme). For comparison, the axis of \Dhb\ is plotted inversely.}
\label{fig:deltar_feii_width}
\end{figure*}

\section{Discussion}
\label{sec:discussion}

\subsection{Shortening of \hb\ Lags in SEAMBHs}

Principal component analysis has reveal that the main variance of quasar
optical spectra is a strong correlation between the flux ratio of broad \feii\
to \hb\ emission, the strength of \oiii$\lambda5007$, and the width of
\hb\ \citep{boroson1992, sulentic2000}.  The so-called Eigenvector 1
has been demonstrated to be driven by Eddington ratio $L_{\rm bol}/L_{\rm
Edd}$ \citep{boroson1992,
sulentic2000, sulentic2000b, marziani2003, shen2014}. AGNs with high Eddington
ratios/accretion rates, such as NLS1s, show
very strong \feii\ emission compared with normal sources \citep{boroson1992,
hu2008a, dong2011}. And at the same time, the \hb\ profiles of NLS1s tend to
be more Lorentzian (smaller values of \Dhb) than those of broader line AGNs,
which probably have more normal accretion rates
\citep{veron2001, zamfir2010, kollatschny2011}. 

\cite{du2016F} recently investigated the correlation between
the strength of \feii, the \hb\ profile, and the accretion rates of the RM AGN
sample. They found that both the relative strength of optical \feii\ lines
($\Rfe=F_{\rm Fe}/F_{\hb}$) and the \hb\ profile shape parameter 
(\Dhb\ = FWHM/$\sigma_{\hb}$) indeed correlate
with accretion rate \dotm, where $F_{\rm Fe}$ is the flux of \feii\ in the 
region 4434--4684 \AA\ and $F_{\hb}$ is the flux of broad \hb. Combining \Dhb\
and \Rfe, they proposed a bivariate correlation of the form $\log \dotm = \alpha_2 + \beta_2
\Dhb + \gamma_2 \Rfe$, termed the ``BLR fundamental plane."
Thus, we can use \Rfe\ and \Dhb\ as proxies of accretion rate to test the 
dependence of $\Delta \rhb$ on accretion rate.

Figure \ref{fig:deltar_feii_width} shows the correlation between $\Delta
\rhb$ and \Rfe, \Dhb, or the \dotm\ derived from the fundamental plane. For
comparison, the axis of \Dhb\ is plotted inversely. In general, the objects
with larger \Rfe\ and smaller \Dhb\ deviate more extremely from the \rl\
relation. \Rfe\ and \Dhb\ are purely observables; thus, using them as proxies
of \dotm\ can avoid the implicit enhancement to the $\Delta \rhb$--\dotm\ correlation
caused by the fact that \dotm\ is a derived quantity.
The scatter of the \dotm--\Rfe\ and \dotm--\Dhb\ correlations is relatively
large (see Figure 1 in \citealt{du2016F}). So, the $\Delta \rhb$--\Rfe\ and
$\Delta \rhb$--\Dhb\ correlations in panels a and b of Figure
\ref{fig:deltar_feii_width} are not as good as the $\Delta \rhb$--\dotm\
correlation in Figure \ref{fig:deltar_mdot}. Panel c in Figure
\ref{fig:deltar_feii_width} shows $\Delta \rhb$ versus \dotm\ derived
from the fundamental plane, which has smaller scatter than each of the single
correlations (\dotm--\Rfe\ or \dotm--\Dhb). 
In consideration of the
relatively large scatter in the \dotm--\Rfe\ and \dotm--\Dhb\ correlations, and
the  fundamental plane, we do not attempt to apply linear regression to Figure
\ref{fig:deltar_feii_width}. They are just used to demonstrate the validity of
Equation (\ref{eqn:deltar_mdot}) in different ways.

\subsection{Self-shadowing Effect of Slim Disks on Lags}

It has been predicted that the self-shadowing effects of slim accretion disks
leads to a shrinking of the ionization front of the BLR, so that the \hb\ lag 
shortens with increasing accretion rate \citep{wang2014c}. Because of 
radiation pressure, the inner part
of the slim disk is not geometrically thin, a property that 
naturally leads to anisotropic ionizing continuum. The vertical thickness of the inner slim disk significantly
suppresses the amount of the ionizing photons that can be received by the BLR
clouds, although it does not change the continuum flux obtained by observers. If
the ionization parameter remains constant for the \hb\ line, the radius at which
\hb\ emits most efficiently will shrink significantly (see more details in
\citealt{wang2014c}). This was first evidenced in the SEAMBH2013--2014 samples (Papers
\citetalias{du2015} and \citetalias{du2016V}). The current SEAMBH2015--2016 sample continues to lend
support to the idea that the shortened \hb\ lags arise from self-shadowing effects of a slim disk.

Self-shadowing effects also depend on BLR geometry. According to the opening
angle of slim disk ($\Delta \Omega_{\rm disk}$), the BLR can be divided into
two parts: 1) a shadowed region and 2) an unshadowed region \citep{wang2014c}. If
BLRs originate from inflows  from the dusty torus to the central BH, as
suggested by \cite{wang2017}, the BLR opening angle ($\Delta\Omega_{\rm BLR}$)
should follow the torus ($\Delta\Omega_{\rm torus}$)\footnote{ It is supported by 
the evidence that $R_{\rm dust}/\rhb$ tends to be a constant for different \dotm\ (see Figure 7 in Paper \citetalias{du2015}), 
where $R_{\rm dust}$ is the inner radius of the dusty torus
measured from the infrared RM observations \citep[e.g.,][]{suganuma2006,
 koshida2014}. It
implies that the BLR and torus have strong connection, and change synchronously.}. Namely,
$\Delta\Omega_{\rm BLR}\sim \Delta\Omega_{\rm torus}$. In the regime of the
standard accretion disk, $\Delta \Omega_{\rm disk}>\Delta \Omega_{\rm BLR}$,
and the entire BLR is illuminated by the radiation of the disk. In the case of
a slim disk, for a simple consideration, the relative  size of the shadowed and
unshadowed regions depends on $\Delta\Omega_{\rm disk}$ and  $\Delta
\Omega_{\rm torus}$. The shadowed region receives less ionizing luminosity 
than the unshadowed region. This may give rise to shrinkage of the ionization
front in the shadowed regions, thereby leading to shortened H$\beta$ lags 
in SEAMBHs.  This is a key prediction of the self-shadowing effects in
SEAMBHs.  Although shortened lags have been observed by the SEAMBH project, 
the relatively longer ones of the unshadowed regions have not yet been 
reported. A possible reason is that our current continuous period of monitoring
is still not long enough because of the regular rainy season from June to
October in the Lijiang site.

We note that the measurement of the \hb\ lag may possibly be influenced by the
characteristic continuum variability timescale \citep[][]{goad2014}.  The
extremely short variability timescale (much shorter than the
intrinsic \hb\ lag or the centroid of the one-dimensional transfer function) may
lead to shortening of the lag measurement \citep[see more detals
in][]{goad2014}. However, the variation timescales of the objects in this
paper are typically much larger than their \hb\ time delays (the
timescales are typically $\sim$100--300 days). Thus, the variation timescale is
unlikely the dominant factor for the shortening of the \hb\ lags.

\subsection{Results from the SDSS-RM campaign}

\begin{figure}
\centering
\includegraphics[width=0.45\textwidth]{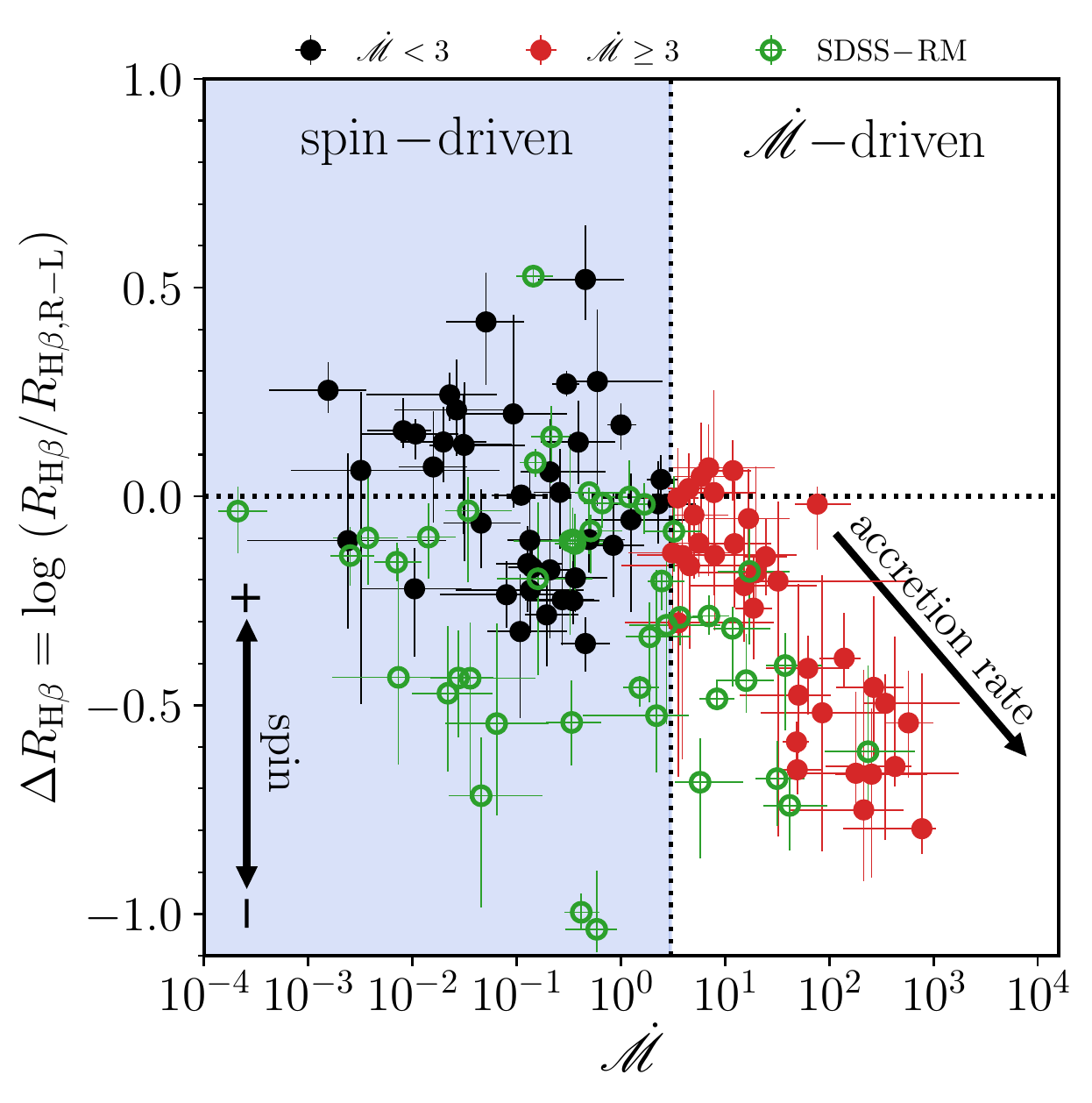} 
\caption{
The $\Delta \rhb - \dotm$ plane, including the SDSS-RM sample. The plane is roughly
divided into two regimes, at $\dotm\approx3$. Low-\dotm\ AGNs have a large scatter in
$\Delta \rhb$, whereas $\Delta \rhb$ is inversely correlated with \dotm\ in the high-\dotm\
regime. The shortening of \hb\ lags in low-\dotm\ AGNs could be explained by retrograde
accretion onto BHs. The arrows
mark the physical drivers for the shortening of \hb\ lags in different accretion regimes.}
\label{fig:mdot_deltar_sdss}
\end{figure}

The large sample size of the Sloan Digital Sky Survey RM (SDSS-RM) project \citep{shen2015} offers a promising opportunity to probe
BHs with a large range of accretion rates
and spins. \cite{grier2017} recently reported \hb\ time lags detected
in the first year of SDSS-RM. They successfully measured \hb\ lags for 44 SDSS 
targets mainly using the Bayesian-based modeling code JAVELIN \citep{zu2011}
and the Continuum REprocessing AGN MCMC software (CREAM;
\citealt{starkey2016}) instead of the traditional ICCF\footnote{Because
they used JAVELIN and CREAM instead of the ICCF to measure the \hb\ time lags, we
do not add them to our analysis in Section \ref{sec:r-l}.}. Interestingly, they, too,
found shortened \hb\ lags compared with the canonical \rl\ 
relation for a number of objects. We plot the 44
objects in the $\Delta \rhb$--\dotm\ plane in Figure
\ref{fig:mdot_deltar_sdss}; the black and red points 
are the same objects in Figure
\ref{fig:deltar_mdot}. The time lags, luminosities, and BH masses are taken
directly from \cite{grier2017}, and their accretion rates $\dotm$
are calculated using Equation (\ref{eqn:mdot}). There are two extreme
cases (SDSS~J142052.44+525622.4 and SDSS~JJ141856.19+535845.0) that show
$\Delta \rhb \approx -1$ and $\dotm \approx 10^{-0.4\sim-0.5}$. A
couple of objects with $\dotm \gtrsim 3$ are in the SEAMBH regime. These 
conform to our expectation 
that high \dotm\ leads to shortened \hb\ lags.  Surprisingly, some
low-\dotm\ objects
{\it also}\ have \hb\ lags much shorter than the \rl\ relation. What is
the physical explanation for these?

One possible interpretation is that this is a signature of retrograde\footnote{``Retrograde'' means the angular momentum of the
BH is opposite to that of the accretion flow.} accretion onto the BH in low-accretion rate AGNs.
In the accretion rate regime of the Shakura-Sunyaev disk ($\dotm \lesssim 3$), 
the inner edge of the disk is fully determined by the last stable radius \citep{Bardeen1972},
since the dissipation of gravitational energy via viscosity can be neglected within this
radius \citep{Page1974}.
Different from the Shakura-Sunyaev disk, the inner edge of a slim disk is mostly determined 
by the accretion rate instead of the spin because the dissipation cannot be neglected \citep{watarai2003}.
As a consequence, except for accretion
rate, the ionizing luminosity highly depends on the spin and hence seriously
influences the ionization front of the BLR \citep{wang2014b}. In the Shakura--Sunyaev
regime, the ionizing luminosity of \hb\ shows a non-monotonic
correlation with the 5100 \AA\ luminosity if the quasar is undergoing
retrograde accretion (see Figure 3 in
\citealt{wang2014b}). A cold disk in retrograde accretion leads to the
inefficient generation of the ionizing continuum, causing the \hb\ region to 
become smaller than the case of prograde accretion \citep{wang2014b}.  For the extreme
case of retrograde accretion onto a  maximally rotating BH, the \hb\ lags are
expected to be shortened by a factor of $\sim 10$ \citep{wang2014b}. 
The two extreme cases from the SDSS-RM campaign may be caused by retrograde
accretion.

In Figure \ref{fig:mdot_deltar_sdss}, we divide the regime of accretion rates by 
$\dotm \approx 3$. \hb\ lags are influenced jointly by the spin and the accretion 
rate below $\dotm \approx 3$, and purely by accretion rate for $\dotm \gtrsim 3$. 
The first regime is spin-driven, the second $\dotm$-driven. The
canonical \rl\, relation only holds for AGNs in the Shakura-Sunyaev regime.
Our proposal should be tested by larger
samples covering the widest possible range of accretion rates. 

The discovery of retrograde accretion onto BHs in AGNs through RM
campaigns, if confirmed by independent measurements (e.g., iron K$\alpha$ observations from X-ray spectra),
has important
implications for the cosmological evolution of BHs. It has been  suggested
that the spin angular momentum of BHs originates from accretion if the 
mass of the BH gained through accretion 
is larger than one-third  of its original mass (e.g.,
\citealt{thorne1974}).  Therefore,  the direction of ongoing accretion may  be
different from the current spins obtained from past accretion episodes. 
The cosmological evolution of the radiation efficiency of  $z\lesssim 2$ quasars \citep{wang2009,li2012} suggests that BHs spinning down with cosmic time.
This is confirmed by simulations of BH evolution \citep{volonteri2013,Tucci2017}. 
The SDSS-RM discovery of shortened lags for AGNs with low accretion rates supports spin-down evolution. 

Additionally, the truncated accretion disk of a BH with low
accretion rate is plausibly responsible for the shortened \hb\ lag. In such
an accretion disk, the linear $L_{5100}-L_{13.6\rm eV}$ relation will be broken
since the ionizing photons are suppressed due to inefficient radiation in the
evaporated part of the advection-dominated accretion flow, where $L_{13.6\rm eV}$
is the ionizing luminosity at 13.6 eV. Details will be presented in a
forthcoming paper (Czerny \& Wang in preparation). 
Broadband spectral energy distributions are needed for these
low-accretion AGNs with shortened lags in order to distinguish the
two possible mechanisms.

\section{Summary}
\label{sec:summary}

We present the observational results of our reverberation mapping campaign of 
super-Eddington accreting massive black holes (SEAMBHs) completed during 2015 -- 2017.
We successfully measured \hb\ time lags for 10 SEAMBHs. Six
 out of the 10
objects have, on average, higher 5100 \AA\ luminosities than the previous
SEAMBHs. The other four targets are found to have \hb\ lags, in general,
consistent with those previously measured during 2013 -- 2015. The new
observations significantly enlarge the size of the SEAMBH RM sample and improve
the completeness of the SEAMBH sample at the high-luminosity end of the \rl\
relation.  The SEAMBH samples show that their \hb\ time lags deviate from the \rl\ relation by
a factor of 2 -- 6 at fixed luminosity. 
The \hb\ lags decreases from the canonical \rl\ relation with 
increasing
\feii/\hb\ flux ratio and change of \hb\ profile. The recent discovery 
by the SDSS-RM collaboration of
\hb\ lags in AGNs with low accretion rates may signify
retrograde accretion onto BHs. This has important 
implications for BH spins. 

\acknowledgments
We acknowledge the support of the staff of the Lijiang
2.4m telescope. Funding for the telescope has been provided
by CAS and the People’s Government of Yunnan
Province. This research is supported by National Key
R\&D Program of China (grants 2016YFA0400701 and 2016YFA0400702), by NSFC through
grants NSFC-11503026, -11173023, -11133006, -11373024, -11233003,
-11473002, -11570326, -11721303, and U1431228, and by Grant
No. QYZDJ-SSW-SLH007 from the Key Research Program
of Frontier Sciences, CAS.

\clearpage
\appendix
\section{Light Curves of Comparison Stars}
\label{sec:comp}
\begin{figure*}
\centering
\includegraphics[width=\textwidth]{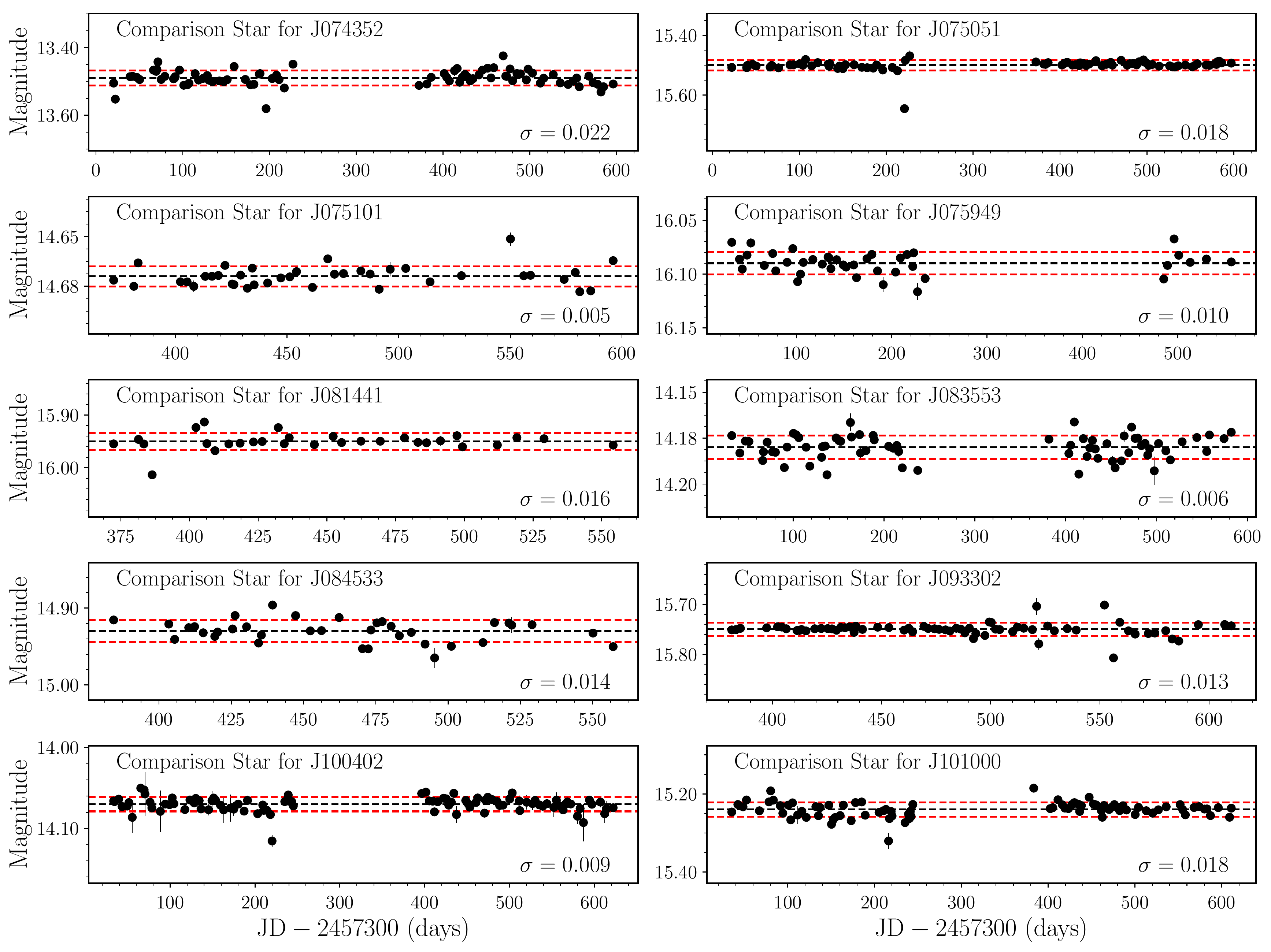} 
\caption{The photometric light curves of the comparison stars in the slit.}
\label{fig:comp}
\end{figure*}

In order to ensure that the comparison stars do not vary significantly during the campaign, we examine
their light curves by performing differential photometry. Several other stars in the same field
were used. The light curves of the comparison stars and their standard deviations are shown in Figure
\ref{fig:comp}. None of the comparison stars shows strong variations during our observation period.  The typical standard deviation is smaller than $\sim2\%$. 

\section{Evaluation to Calibration Precision}
\label{sec:calib}

\begin{figure*}
\centering
\includegraphics[width=0.65\textwidth]{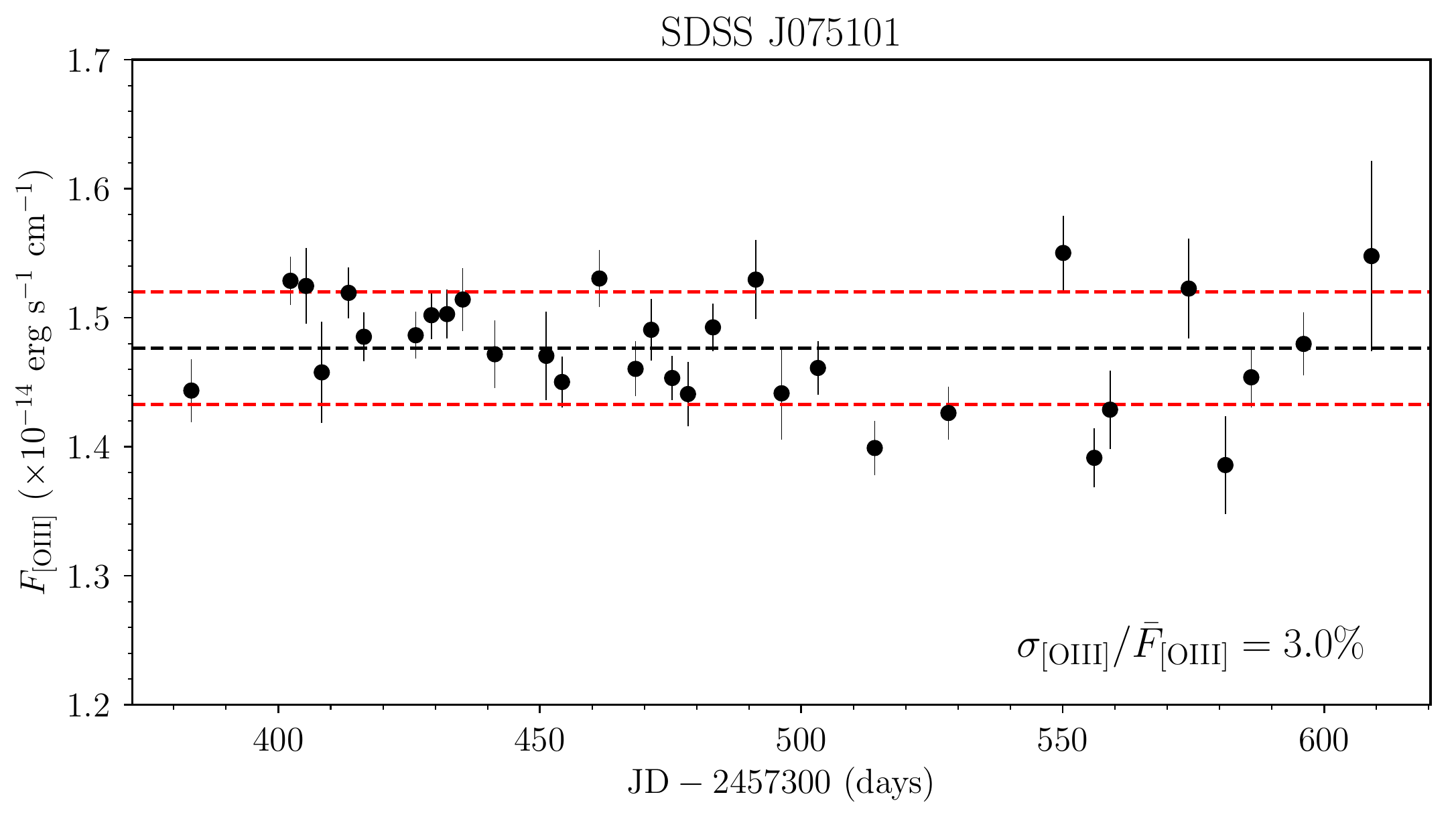}
\caption{\oiii\ light curve of SDSS J075101. The black dashed line marks the average value of \oiii\ fluxes, and 
the red dashed lines mark the $\pm1\sigma$ standard deviation. The ratio between the standard deviation and the mean value
is 3.0\%.}
\label{fig:oiii_075101}
\end{figure*}

To show the precision of the comparison-star calibration,
we plot the \oiii\ light curve of SDSS J075101 after the calibration. This
object is the one that has the strongest \oiii\ in the sample; it has 
relatively weak \feii\ and high S/N. Its \oiii\ fluxes are
measured by a simple multi-component fitting (similar to Paper
\citetalias{hu2015}): we model the broad \hb\ line with two Gaussians, each of the
other lines (\oiii\ $\lambda\lambda4959,5007$, narrow \hb, and \heii) with one Gaussian,  the
\feii\ using the template from \cite{boroson1992}, and the continuum with a power
law.  The scatter of its \oiii\ flux is 3.0\% (Figure \ref{fig:oiii_075101}), which can be regarded as an estimate of the
calibration precision. However, it should be noted that the fitting itself may
introduce large uncertainty to the \oiii\ flux measurement, because \oiii\ is too weak and \feii\ is relatively strong
in SEAMBHs. Thus, the value of 3.0\% is only an upper limit on the
calibration uncertainty. For the other objects with weaker \oiii, 
stronger \feii, and lower S/N, it is difficult to obtain reliable \oiii\ light curves.

\section{Mean and RMS spectra}
\label{sec:meanrms}

\begin{figure*}
\includegraphics[width=\textwidth]{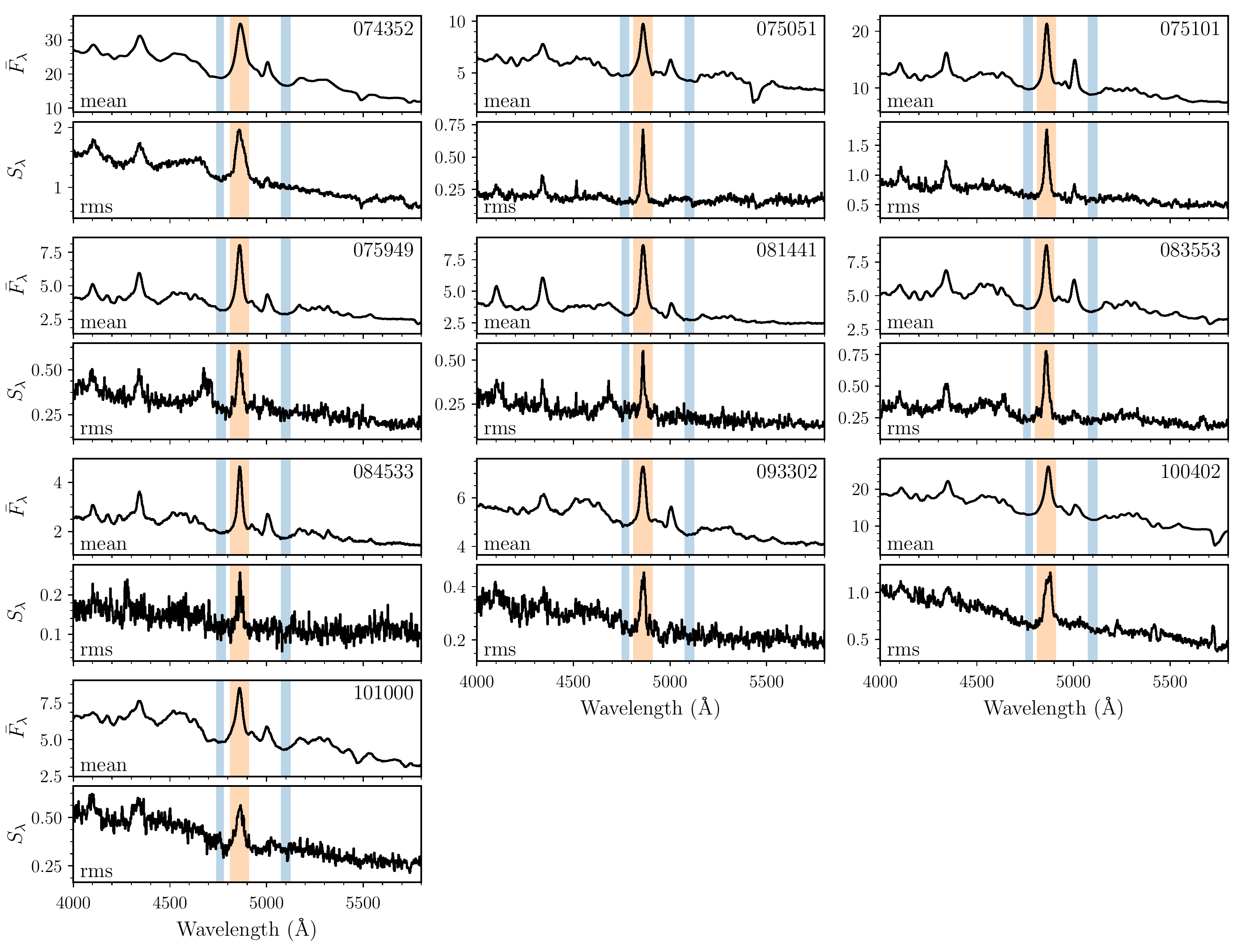}
\caption{Mean and RMS spectra (observed flux vs. rest-frame wavelength) of the objects. The orange and 
blue regions are the windows for \hb\ emission lines and their backgrounds. The unit is ${10^{-16}\ \rm erg\ s^{-1}\ cm^{-2}\ \AA^{-1}}$. }
\label{fig:meanrms}
\end{figure*}

To illustrate the general spectral characteristics of each object, we plot their  
mean spectra and root-mean-square (RMS) spectra in Figure \ref{fig:meanrms}. Following 
the procedures in Papers \citetalias{du2015} and \citetalias{du2016V}, the mean and RMS spectra 
are defined, respectively, as
\begin{equation}
\bar{F}_{\lambda}=\frac{1}{N}\sum_{i=1}^NF_{\lambda}^i,
\end{equation}
and 
\begin{equation}
S_{\lambda}=\left[\frac{1}{N}\sum_{i=1}^N\left(F_{\lambda}^i-\bar{F}_{\lambda}\right)^2\right]^{1/2}.
\end{equation}
$F_{\lambda}^i$ is the $i$-th spectrum of the object, and $N$ is the number of spectra it has. 
It is obvious that their \feii\ emission lines are strong and \oiii\ lines are extremely weak, which are 
the typical characteristics of AGNs with high accretion rates \citep[see, e.g.,][]{boroson1992, shen2014}.

\begin{thebibliography}{}
   \bibitem[Bahcall et al.(1972)]{bahcall1972} Bahcall, J. N., Kozlovsky, B.-Z., \& Salpeter, E. E. 1972, \apj, 171, 467
   \bibitem[Bardeen et al.(1972)]{Bardeen1972} Bardeen, J. M., Press, W. H., Teukolsky, S. A. 1972, \apj, 178, 347
   \bibitem[Barth et al.(2015)]{barth2015} Barth, A. J., Bennert, V. N., Canalizo, G., et al. 2015, \apjs, 217, 26
   \bibitem[Barth et al.(2013)]{barth2013} Barth, A. J., Pancoast, A., Bennert, V. N., et al. 2013, \apj, 769, 128
   \bibitem[Barth et al.(2011)]{barth2011} Barth. A. J., Pancoast, A., Thorman, S. J., et al. 2011, \apj, 743, 4
   \bibitem[Beloborodov(1998)]{beloborodov1998} Beloborodov, A.~M.\ 1998, \mnras, 297, 739
   \bibitem[Bentz et al.(2016)]{bentz2016ugc} Bentz, M.~C., Batiste, M., Seals, J., et al.\ 2016, \apj, 831, 2
   \bibitem[Bentz et al.(2016)]{bentz2016mcg} Bentz, M.~C., Cackett, E.~M., Crenshaw, D.~M., et al.\ 2016, \apj, 830, 136
   \bibitem[Bentz et al.(2013)]{bentz2013} Bentz, M. C., Denney, K. D., Grier, C. J., et al. 2013, \apj, 767, 149
   \bibitem[Bentz et al.(2008)]{bentz2008} Bentz, M. C., Walsh, J. L., Barth, A. J., et al. 2008, \apj, 689, L21
   \bibitem[Bentz et al.(2009)]{bentz2009} Bentz, M. C., Walsh, J. L., Barth, A. J., et al. 2009, \apj, 705, 199
   \bibitem[Blandford \& McKee(1982)]{blandford1982} Blandford, R.~D., \& McKee, C.~F.\ 1982, \apj, 255, 419
   \bibitem[Boroson \& Green(1992)]{boroson1992} Boroson, T.~A., \& Green, R.~F.\ 1992, \apjs, 80, 109
   \bibitem[Denney et al.(2009)]{denney2009} Denney, K. D., Peterson, B. M., Pogge, R. W., et al. 2009, \apj, 704, L80
   \bibitem[Denney et al.(2010)]{denney2010} Denney, K.~D., Peterson, B.~M., Pogge, R.~W., et al.\ 2010, \apj, 721, 715
   \bibitem[Dong et al.(2011)]{dong2011} Dong, X.-B., Wang, J.-G., Ho, L.~C., et al.\ 2011, \apj, 736, 86
   \bibitem[Du et al.(2014)]{du2014} Du, P., Hu, C., Lu, K.-X., et al. (SEAMBH Collaboration) 2014, \apj, 782, 45 (Paper I)
   \bibitem[Du et al.(2015)]{du2015} Du, P., Hu, C., Lu, K.-X., et al. (SEAMBH Collaboration) 2015, \apj, 806, 22 (Paper IV)
   \bibitem[Du et al.(2016a)]{du2016VI} Du, P., Lu, K.-X., Hu, C., et al. (SEAMBH Collaboration) 2016a, \apj, 820, 27 (Paper VI)
   \bibitem[Du et al.(2016b)]{du2016V} Du, P., Lu, K.-X., Zhang, Z.-X., et al. (SEAMBH Collaboration) 2016b, \apj, 825, 126  (Paper V)
   \bibitem[Du et al.(2016c)]{du2016F} Du, P., Wang, J.-M., Hu, C., et al.\ 2016c, \apjl, 818, L14
   \bibitem[Du et al.(2017)]{du2017} Du, P., Wang, J.-M., \& Zhang, Z.-X.\ 2017, \apjl, 840, L6 
   \bibitem[Fausnaugh(2017)]{fausnaugh2017a} Fausnaugh, M.~M.\ 2017, \pasp, 129, 024007
   \bibitem[Fausnaugh et al.(2017)]{fausnaugh2017} Fausnaugh, M.~M., Grier, C.~J., Bentz, M.~C., et al.\ 2017, \apj, 840, 97
   \bibitem[Fischer et al.(2014)]{fischer2014} Fischer, T.~C., Crenshaw, D.~M., Kraemer, S.~B., Schmitt, H.~R., \& Turner, T.~J.\ 2014, \apj, 785, 25
   \bibitem[Frank et al.(2002)]{frank2002} Frank, J., King, A., \& Raine, D.~J.\ 2002, Accretion Power in Astrophysics, by Juhan Frank and Andrew King and Derek Raine, pp.~398.~ISBN 0521620538.~Cambridge, UK: Cambridge University Press, February 2002., 398
   \bibitem[Gaskell \& Peterson(1987)]{gaskell1987} Gaskell, C.~M., \& Peterson, B.~M.\ 1987, \apjs, 65, 1 
   \bibitem[Gaskell \& Sparke(1986)]{gaskell1986} Gaskell, C.~M., \& Sparke, L.~S.\ 1986, \apj, 305, 175
   \bibitem[Goad \& Korista(2014)]{goad2014} Goad, M.~R., \& Korista, K.~T.\ 2014, \mnras, 444, 43
   \bibitem[Grier et al.(2013)]{grier2013} Grier, C.~J., Martini, P., Watson, L.~C., et al.\ 2013, \apj, 773, 90
   \bibitem[Grier et al.(2012)]{grier2012} Grier, C.~J., Peterson, B.~M., Pogge, R.~W., et al.\ 2012, \apj, 755, 60 
   \bibitem[Grier et al.(2017)]{grier2017} Grier, C.~J., Trump, J.~R., Shen, Y., et al.\ 2017, arXiv:1711.03114 
   \bibitem[Ho \& Kim(2014)]{ho2014} Ho, L.~C., \& Kim, M.\ 2014, \apj, 789, 17 
   \bibitem[Ho \& Kim(2015)]{ho2015} Ho, L.~C., \& Kim, M.\ 2015, \apj, 809, 123
   \bibitem[Horne(1994)]{horne1994} Horne, K.\ 1994, Reverberation Mapping of the Broad-Line Region in Active Galactic Nuclei, 69, 23
   \bibitem[Hu et al.(2015)]{hu2015} Hu, C., Du, P., Lu, K.-X., et al. (SEAMBH Collaboration) 2015, \apj, 804, 138 (Paper III)
   \bibitem[Hu et al.(2008a)]{hu2008a} Hu, C., Wang, J.-M., Ho, L.~C., et al.\ 2008a, \apj, 687, 78-96
   \bibitem[Hu et al.(2008b)]{hu2008b} Hu, C., Wang, J.-M., Ho, L.~C., et al.\ 2008b, \apjl, 683, L115
   \bibitem[Hu et al.(2016)]{hu2016} Hu, C., Wang, J.-M., Ho, L.~C., et al.\ 2016, \apj, 832, 197
   \bibitem[Jiang et al.(2016)]{jiang2016} Jiang, L., Shen, Y., McGreer, I.~D., et al.\ 2016, \apj, 818, 137
   \bibitem[Kaspi et al.(2007)]{kaspi2007} Kaspi, S., Brandt, W. N., Maoz, D., et al. 2007, \apj, 659, 997
   \bibitem[Kaspi et al.(2005)]{kaspi2005} Kaspi, S., Maoz, D., Netzer, H., et al.\ 2005, \apj, 629, 61 
   \bibitem[Kaspi et al.(2000)]{kaspi2000} Kaspi, S., Smith, P. S., Netzer, H., et al. 2000, \apj, 533, 631
   \bibitem[Kollatschny \& Zetzl(2011)]{kollatschny2011} Kollatschny, W., \& Zetzl, M.\ 2011, \nat, 470, 366
   \bibitem[Koshida et al.(2014)]{koshida2014} Koshida, S., Minezaki, T., Yoshii, Y., et al.\ 2014, \apj, 788, 159
   \bibitem[Kewley et al.(2006)]{kewley2006} Kewley, L.~J., Groves, B., Kauffmann, G., \& Heckman, T.\ 2006, \mnras, 372, 961
   \bibitem[Laor \& Netzer(1989)]{laor1989} Laor, A., \& Netzer, H.\ 1989, \mnras, 238, 897
   \bibitem[Li et al.(2012)]{li2012} Li, Y.-R., Wang, J.-M., \& Ho, L.~C.\ 2012, \apj, 749, 187
   \bibitem[Li et al.(2018)]{li2018} Li, Y.-R., Songsheng, Y.-Y., Qiu, J., et al. (SEAMBH Collaboration) 2018, \apj, submitted (Paper VIII)
   \bibitem[Lu et al.(2016)]{lu2016} Lu, K.-X., Du, P., Hu, C., et al.\ 2016, \apj, 827, 118
   \bibitem[Marziani et al.(2003)]{marziani2003} Marziani, P., Zamanov, R.~K., Sulentic, J.~W., \& Calvani, M.\ 2003, \mnras, 345, 1133 
   \bibitem[Mathur et al.(2012)]{mathur2012} Mathur, S., Fields, D., Peterson, B.~M., \& Grupe, D.\ 2012, \apj, 754, 146 
   \bibitem[McLure \& Dunlop(2002)]{mclure2002} McLure, R.~J., \& Dunlop, J.~S.\ 2002, \mnras, 331, 795
   \bibitem[Onken et al.(2004)]{onken2004} Onken, C.~A., Ferrarese, L., Merritt, D., et al.\ 2004, \apj, 615, 645 
   \bibitem[Page \& Thorne(1974)]{Page1974}Page, D. N. \& Thorne, K. S. 1974, \apj, 191, 499
   \bibitem[Pancoast et al.(2014)]{pancoast2014} Pancoast, A., Brewer, B.~J., Treu, T., et al.\ 2014, \mnras, 445, 3073
   \bibitem[Park et al.(2012)]{park2012} Park, D., Kelly, B.~C., Woo, J.-H., \& Treu, T.\ 2012, \apjs, 203, 6
   \bibitem[Peterson et al.(1993)]{peterson1993} Peterson, B. M., Ali, B., Horne, K., et al. 1993, \pasp, 105, 247
   \bibitem[Peterson et al.(2002)]{peterson2002} Peterson, B. M., Berlind, P., Bertram, R., et al. 2002, \apj, 581, 197
   \bibitem[Peterson et al.(2004)]{peterson2004} Peterson, B. M., Ferrarese, L., Gilbert, K. M., et al. 2004, \apj, 613, 682
   \bibitem[Peterson et al.(1998)]{peterson1998} Peterson, B. M., Wanders, I., Bertram, R., et al. 1998, \apj, 501, 82
   \bibitem[Planck Collaboration et al.(2014)]{ade2014} Planck Collaboration, Ade, P.~A.~R., Aghanim, N., et al.\ 2014, \aap, 571, A16 
   \bibitem[Rafter et al.(2011)]{rafter2011} Rafter, S. E., Kaspi, S., Behar, E., Kollatschny, W., \& Zetzl, M. 2011, \apj, 741, 66
   \bibitem[Rafter et al.(2013)]{rafter2013} Rafter, S. E., Kaspi, S., Chelouche, D., et al. 2013, \apj, 773, 24
   \bibitem[S{\c a}dowski et al.(2011)]{sadowski2011} S{\c a}dowski, A., Abramowicz, M., Bursa, M., et al.\ 2011, \aap, 527, A17 
   \bibitem[Schlafly \& Finkbeiner(2011)]{schlafly2011} Schlafly, E.~F., \& Finkbeiner, D.~P.\ 2011, \apj, 737, 103
   \bibitem[Shakura \& Sunyaev(1973)]{shakura1973} Shakura, N.~I., \& Sunyaev, R.~A.\ 1973, \aap, 24, 337
   \bibitem[Shen et al.(2015)]{shen2015} Shen, Y., Brandt, W.~N., Dawson, K.~S., et al.\ 2015, \apjs, 216, 4 
   \bibitem[Shen \& Ho(2014)]{shen2014} Shen, Y., \& Ho, L.~C.\ 2014, \nat, 513, 210
   \bibitem[Shen et al.(2016)]{shen2016} Shen, Y., Horne, K., Grier, C.~J., et al.\ 2016, \apj, 818, 30 
   \bibitem[Shen et al.(2011)]{shen2011} Shen, Y., Richards, G.~T., Strauss, M.~A., et al.\ 2011, \apjs, 194, 45
   \bibitem[Starkey et al.(2016)]{starkey2016} Starkey, D.~A., Horne, K., \& Villforth, C.\ 2016, \mnras, 456, 1960
   \bibitem[Stern \& Laor(2013)]{stern2013} Stern, J., \& Laor, A.\ 2013, \mnras, 431, 836
   \bibitem[Suganuma et al.(2006)]{suganuma2006} Suganuma, M., Yoshii, Y., Kobayashi, Y., et al.\ 2006, \apj, 639, 46
   \bibitem[Sulentic et al.(2000a)]{sulentic2000} Sulentic, J.~W., Marziani, P., \& Dultzin-Hacyan, D.\ 2000, \araa, 38, 521 
   \bibitem[Sulentic et al.(2000b)]{sulentic2000b} Sulentic, J.~W., Zwitter, T., Marziani, P., \& Dultzin-Hacyan, D.\ 2000, \apjl, 536, L5
   \bibitem[Thorne(1974)]{thorne1974} Thorne, K.~S.\ 1974, \apj, 191, 507 
   \bibitem[Tremaine et al.(2002)]{tremaine2002} Tremaine, S., Gebhardt, K., Bender, R., et al.\ 2002, \apj, 574, 740
   \bibitem[Tucci \& Volonteri(2017)]{Tucci2017} Tucci, M. \& Volonteri, M. 2017, \aap, 600, 64
   \bibitem[van Groningen \& Wanders(1992)]{vanGroningen1992} van Groningen, E., \& Wanders, I.\ 1992, \pasp, 104, 700
   \bibitem[Volonteri et al.(2013)]{volonteri2013} Volonteri, M., Sikora, M., Lasota, J.-P. \& Merloni, A.\ 2013, \apj, 775, 94
   \bibitem[V{\'e}ron-Cetty et al.(2001)]{veron2001} V{\'e}ron-Cetty, M.-P., V{\'e}ron, P., \& Gon{\c c}alves, A.~C.\ 2001, \aap, 372, 730
   \bibitem[Vestergaard \& Peterson(2006)]{vestergaard2006} Vestergaard, M., \& Peterson, B.~M.\ 2006, \apj, 641, 689
   \bibitem[Wang et al.(2017)]{wang2017} Wang, J.-M., Du, P., Brotherton, M.~S., et al.\ 2017, Nature Astronomy, 1, 775
   \bibitem[Wang et al.(2014a)]{wang2014} Wang, J.-M., Du, P., Hu, C., et al. (SEAMBH Collaboration) 2014a, \apj, 793, 108 (Paper II)
   \bibitem[Wang et al.(2014b)]{wang2014b} Wang, J.-M., Du, P., Li, Y.-R., et al.\ 2014b, \apjl, 792, L13
   \bibitem[Wang et al.(2013)]{Wang2013}Wang, J.-M., Du, P., Valls-Gabaud, D., Hu, C. \& Netzer, H. 2013, \prl, 110, 081301   
   \bibitem[Wang et al.(2009)]{wang2009} Wang, J.-M., Hu, C., Li, Y.-R., et al.\ 2009, \apjl, 697, L141 
   \bibitem[Wang et al.(2014c)]{wang2014c} Wang, J.-M., Qiu, J., Du, P., \& Ho, L.~C.\ 2014c, \apj, 797, 65
   \bibitem[Watarai \& Mineshige(2003)]{watarai2003} Watarai, K.-Y. \& Mineshige, S. 2003, \pasj, 55, 959
   \bibitem[Woo et al.(2010)]{woo2010} Woo, J.-H., Treu, T., Barth, A.~J., et al.\ 2010, \apj, 716, 269
   \bibitem[Woo et al.(2015)]{woo2015} Woo, J.-H., Yoon, Y., Park, S., Park, D., \& Kim, S.~C.\ 2015, \apj, 801, 38
   \bibitem[Xiao et al.(2018)]{xiao2018} Xiao, M., Du, P., Horne, K., et al. (SEAMBH Collaboration) 2018, \apj, submitted (Paper VII)
   \bibitem[Zamfir et al.(2010)]{zamfir2010} Zamfir, S., Sulentic, J.~W., Marziani, P., \& Dultzin, D.\ 2010, \mnras, 403, 1759
   \bibitem[Zu et al.(2011)]{zu2011} Zu, Y., Kochanek, C.~S., \& Peterson, B.~M.\ 2011, \apj, 735, 80
\end{thebibliography}
\end{document}